\newcommand{\Msun}[0]{\mbox{ M}_\odot}

\newcommand{\Zdot}[0]{\mbox{ Z}_\odot}
\newcommand{\Zsun}[0]{\mbox{ Z}_\odot}
\newcommand{\Lya}[0]{Ly$\alpha$}
\newcommand{\Flux}[0]{$\times 10^{-17}\mbox{ erg s}^{-1}\mbox{ cm}^{-2}\mbox{ }$}
\newcommand{\cms}[0]{\mbox{ cm s}^{-1}}
\newcommand{\Hb}[0]{H$\beta$}

\newcommand{\CosmosI}[0]{LAE40844}
\newcommand{\CosmosII}[0]{LAE27878}

\newcommand{\etal}[0]{et al.}

\newcommand{\fig}[1]{Figure~\ref{#1}}		
\newcommand{\sect}[1]{\S\ref{#1}}			
\newcommand{\tabl}[1]{Table~\ref{#1}}		

\documentclass{emulateapj}
\slugcomment{Submitted to AJ}

\begin{document}



\title{High-Redshift Galaxies with Large Ionization Parameters}

\author{Mark L. A. Richardson\altaffilmark{1}, Emily M. Levesque\altaffilmark{2}, Emily M. McLinden\altaffilmark{1,}\altaffilmark{3}, Sangeeta Malhotra\altaffilmark{1}, James E. Rhoads\altaffilmark{1}, Lifang Xia\altaffilmark{1}}
\altaffiltext{1}{School of Earth and Space Exploration, Arizona State University, Tempe, AZ 85287}
\altaffiltext{2}{CASA, Department of Astrophysical and Planetary Sciences, University of Colorado 389-UCB, Boulder, CO 80309}
\altaffiltext{3}{George P. and Cynthia Woods Mitchell Institute for Fundamental Physics and Astronomy, Department of Physics and Astronomy, Texas A\&M University}
\shortauthors{Richardson \emph{et al.}} 

\setcounter{footnote}{3}


\begin{abstract}
Motivated by recent observations of galaxies dominated by emission lines, which show evidence of being metal poor with young stellar populations, we present calculations of multiple model grids with a range of abundances, ionization parameters, and stellar ages, finding that the predicted spectral line diagnostics are heavily dependent on all three parameters. These new model grids extend the ionization parameter to larger values than typically explored. We compare these model predictions with previous observations of such objects, including two new Lyman-$\alpha$ emitting galaxies (LAE) that we have observed. Our models give improved constraints on the metallicity and ionization parameter of these previously studied objects, as we are now able to consider high ionization parameter models. However, similar to previous work, these models have difficulty predicting large line diagnostics for high ionization 
potential species, requiring future work refining the modelling of FUV photons. Our model grids are also able to constrain the metallicity and ionization parameter of our LAEs, and give constraints on their Ly$\alpha$ escape fractions, all of which are consistent with recent lower redshift studies of LAEs.
\end{abstract}


\maketitle

\section{Introduction} \label{intro}
%
Understanding the history of galactic evolution requires us to probe the early universe when the first
star-formation episodes were occurring. It is during these events that the first metals are created, via large, 
hot, young stellar populations. These hot stars drive efficient ionization, which is quantified by the ionization 
parameter, $q\equiv S_0/n_H$, the ratio of the mean ionizing photon flux, $S_0$, and the mean atom density, 
$n_H$ \citep{Dopita06}. These hot stars have short lives, thus the population's flux is strongly dependent on 
stellar age.

%
The metals produced by massive stars allow for more efficient cooling in nebular gas, leading to the efficient  formation of enriched stars. 
A galaxy's star-formation history can therefore be assessed through its metallicity \citep{Tinsley80}. The repeating cycle of
star formation and evolution enriches the interstellar medium with the products 
of nuclear burning from each generation of stars. The metallicity of a galaxy is extremely important for setting the 
temperature of stars in stellar synthesis models, for accurately predicting equivalent widths, as well as 
illustrating the chemical enrichment history of the galaxy. Assuming very little metals accumulate from 
accretion from the intergalactic medium (IGM) then the metallicity of the galaxy should be a lower limit 
for all of the metals created in past star-forming cycles. It is merely a lower limit as some of these metals 
have been ejected in energetic outflows from supernovae and stellar winds to enrich the IGM further, while 
accretion of metal-poor gas will lower the observed metal abundance.

Previous work has demonstrated that the metallicity is correlated with the stellar mass of the galaxy 
(Zaritsky \etal\ 1994; Tremonti \etal\ 2004; Erb \etal\ 2006; Mannucci \etal\ 2010; Xia \etal\ 2012). Among the youngest and least massive galaxies (see Finkelstein \etal\ 2011 and 
references therein) are Lyman-$\alpha$ emitters (LAE). These galaxies are host to young 
stellar populations and have a significant number of hot, 
massive stars whose ionizing radiation will mainly escape from the galaxy as Ly$\alpha$ 
photons. The \Lya\ line itself can contain up to 6\% of the total emitted light from a star-forming galaxy 
\citep{Partridge67}, yielding the location of the galaxy and information on its star-formation history.
While their specific star formation is quite large, LAEs typically have an absolute star 
formation an order of magnitude smaller than other high-redshift galaxies, such as Lyman-break galaxies, 
and are thus an excellent test of the low mass, low star-formation regime of 
the fundamental metallicity relation (FMR; Mannucci \etal\ 2010), which was in good agreement for redshifts $z\leq
2.5$, but inconsistent with other $z\sim3$ galaxies.

%
The ionization parameter of a galaxy is suggestive of the effective temperature of its stellar population. Larger ionization parameters than $2\times 10^8$ cm s$^{-1}$, although uncommon in the local universe (See Rigby \& Rieke 2004; Lilly \etal\ 2003), may very well be present in the early universe around stronger, younger starbursts. Indeed, earlier work by \citet{Fosbury03}, \citet{Richard11}, and \citet{Erb10} and very recent work by \citet{Nakajima13}, \citet{Jaskot13} and \citet{Xia12} have demonstrated that these earlier galaxies undergoing significant starburst do have larger ionization parameters. These same objects also
have low metallicities, as expected. 

%
Large ionization parameter galaxies will excite more nebular emission lines, which act to cool the gas. Thus to maintain
the observed strong emission lines, there must be a maintained source of ionizing flux. This suggests either a galaxy 
with a near constant rate of star formation, where new bright hot stars are replacing 
the recently deceased stars, or a very recent,
nearly instantaneous epoch of star formation, which is transient and unable to maintain a high ionization for more
than a few Myr. While a continuous star formation will reach equilibrium and then be age-independent \citep{Kewley01}, a stellar population
that was created nearly instantaneously will begin losing the very brightest stars after a very short amount of time, and the O-B population as a whole and the temperature of its nebular gas will diminish significantly by about 5 Myr (Schaller \etal\ 1992; 
Schaerer \etal\ 1993; Charbonnel \etal\ 1993; Levesque \etal\ 2010). Left undisturbed by feedback effects, the most metal rich nebular gas will cool efficiently 
such that beyond this age almost no emission lines will be visible. 

%
With suitably chosen emission lines, we can constrain the metallicity and
ionization parameter. For the redshifts and fluxes of the 
objects we study in this work, we are limited to observing [OII] 3727 \AA, [NeIII] 3869 \AA\, the [OIII] 4959,5007 \AA\  doublet, and H$\beta$ 4861 \AA. Thus we can compare our observations with the [OIII] 5007 \AA/[OII] and [OIII] 5007 \AA/H$\beta$ line diagnostics, both of which
were described in \citet{Levesque10}. Although these line diagnostics are highly dependent on ionization
parameter, they have been shown (e.g., Kewley \etal\ 2002) to depend on metallicity as well. 
Note that there exists a degenerate dependence of the [OIII]/H$\beta$ diagnostic on the metallicity of the galaxy, 
making it a difficult diagnostic to use without other constraints. This degeneracy stems from the fact that 
at low abundances the amount of ionized oxygen grows with the 
metallicity until reaching a critical abundance where the photoionized region can efficiently cool, above which 
the ionized oxygen then drops with increasing 
abundance \citep{Kewley02}. Typically this degeneracy is broken by observing another line diagnostic, 
such as [NII]/[OII]. However, as we will demonstrate, if the [OIII]/[OII] ratio, which is monotonic in both metallicity and ionization
parameter, predicts a sufficiently 
low metallicity and high ionization parameter, then this ratio will also break the degeneracy of the [OIII]/H$\beta$ diagnostic.

The stellar age can be well constrained by observing the continuum flux in several spectral bands. By combining
these observations with stellar synthesis models, one can determine the best fit models which predict a stellar formation
history and age (\emph{e.g.} Yan \etal\ 2005). This method probes the total population of stars, instead of just the hottest stars
that are  probed by observing
emission lines.

%
Thus, whether one is constraining the stellar age, or the metallicity and ionization parameter, one must compare
observations with models. \citet{Levesque10} combined the output of the stellar population synthesis code Starburst99  (Leitherer et al. 1999; V\'{a}zquez \& 
Leitherer 2005) with the photoionization code 
Mappings III (Binette et al. 1985; Sutherland \& Dopita 1993) to model the emission line diagnostics capable of constraining the metallicity and 
ionization parameter given a star formation history and age.
Their work illustrated how different stellar population formation histories and ages, metallicities, densities, stellar mass-loss models, and ionization parameters can vastly change the ratios between different spectral lines. Such a study is 
essential for attempting to probe the conditions of galaxies at all redshifts. However, their work focused on ionization
parameters more typical of low-redshift, older galaxies, with values of $q \le 4\times 10^8$ cm s$^{-1}$.

%
Earlier work, including \citet{Kewley02}, only considered one stellar age and formation history. \citet{Levesque10} have demonstrated that this is insufficient, as older, high metallicity populations 
are typically too cold to match observed emission line strengths. Recent observations
that demonstrate ionization parameters beyond those studied by \citet{Kewley02} are still using this work on which to base their conclusions. This demonstrates the need to continue the work of \citet{Levesque10} for the highest
ionization parameters, allowing for more detailed study of the new population of young high redshift galaxies.

In this work, we have observed two high redshift galaxies having bright \Lya\ lines, 
and strong [OIII] 5007 \AA\ detection. One has its ionization parameter well matched with the upper end of the work of \citet{Levesque10}, while the second has its ionization parameter above their work. There is a clear gap in models for higher ionization parameter, thus we continue their work for the higher ionization parameter values. We use these models to better
constrain our larger ionization parameter object, and apply them to other large ionization parameter objects in Appendix A.

The structure of this paper is as follows. We begin by discussing our sample, observations and data reduction 
in \sect{sample}. We then present the measured line emission in \sect{reduction}, and discuss motivation for requiring higher ionization parameter models. In \sect{models} we introduce the ionization models created using MAPPINGS III, presenting multiple line diagnostics relevant to this work. In \sect{results} we apply our models to our sample to determine what constraints they place on the ionization parameter, metallicity, H$\beta$ 
flux and \Lya\ escape fraction. We conclude in \sect{conclusion}. Additional line diagnostic comparisons and applications with multiple galaxies in the literature are presented 
in Appendix A.
%
%
%
%
%
\section{Observations} \label{sample}
We observed two \Lya\ emitting galaxies, 
\CosmosI\ and \CosmosII, whose selection is described in \citet{McLinden11}.   \CosmosI\ was previously 
detected in \Lya\ with a redshift of $z_{Ly\alpha}=3.11639 \pm 0.00021$, and \CosmosII\ was previously 
detecting in \Lya\ with a redshift of $z_{Ly\alpha}=3.12051 \pm 0.00021$. We performed subsequent IR spectroscopic 
observations at Gemini North with the Near-infrared Integral Field Spectrometer (NIFS; 
McGregor \etal\ 2003) which uses an Integral Field Unit (IFU) with a field-of-view of 2.99" $\times$ 2.97" 
over 29 $\times$ 69 pixels. This device has a spectral resolving power of roughly 9600 in the two bands 
we used.  Observations of \CosmosI\ for [OII] and [OIII] using H-Band (1.65 $\mu$m) and K-Band (2.20 $\mu$m) 
were done on March 4$^{th}$, 2010, and observations of \CosmosII\ for [OII], H$\beta$, and [OIII] using H-Band 
(1.65 $\mu$m) and K-Band (2.00 $\mu$m and 2.20 $\mu$m) were done on March 5$^{th}$ and April 27
$^{th}$, 2010. We did not observe H$\beta$ for \CosmosI\  as its systemic redshift, as determined by 
\citet{McLinden11}, placed this line in a region of strong telluric absorption.  Based on our limited 
observing time we settled on solely observing H$\beta$ for \CosmosII.
We note that on all nights our objects were unresolved in our $\sim$ 0.6" seeing,
 consistent with a LAE galaxy of size $\sim1$ kpc (see Malhotra \etal\ 2012) at an angular-size 
distance of 1600 Mpc at z$\simeq$3.1.

Each science object was observed for 300 sec per exposure with 30 (13) exposures in the K-
Band, centred on 2.20$\mu \mbox{m}$, and 20 (11) exposures in the H-Band, centred on 1.65$\mu \mbox{m}$, for \CosmosI\ (\CosmosII). 
\CosmosII\ also had 16 exposures in K-Band, centred on 2.00 $\mu \mbox{m}$, to look at H$\beta$.
Each science object was coupled with 30 sec exposures of two A0V standard stars to later correct 
for telluric absorption and to perform the flux calibration. Finally, calibration frames were taken at the end 
of the observing night, including darks, flats and spectral correction frames to rectify the image plane 
into the IFU.

The reduction process almost completely followed the reduction tasks provided by Gemini. This pipeline 
was designed to develop the datacube and included flat-field and dark correction as well as some sky 
line subtraction and telluric absorption correction. Of this, we modified the sky line subtraction and 
postponed the telluric absorption correction. 

Our observations did not include sky frames, thus to compose a sky frame for each science we took the 
previous two exposures and following two exposures and median-combined them. As the offset pattern
had five unique positions, this allowed only the sky to be preserved. For the first and last (second 
and second-to-last) exposures, the sky was created by median combining the three succeeding  and 
three preceding (preceding plus three succeeding and three preceding plus succeeding) exposures, 
respectively. A second sky-subtraction was performed by determining the median value of the 
non-science area of the final datacube at each spectral plane and subtracting that from the entire plane. 
This method removed particular sky fluctuations specific to the exposure.

The telluric absorption correction was performed simultaneously with the flux calibration, which was 
completed by using the Pickles Model Spectra \citep{Pickles98} for A0V stars. \CosmosI\ was 
calibrated by scaling the A0V model spectra to match the observed H- and K-Band 2MASS magnitudes 
\citep{Cutri03} of HIP41798 and HIP73200. Once scaled, this spectrum was divided by the observed 
spectrum to get a sensitivity function relating counts to $F_\lambda$. This simulataneously corrected for telluric 
absorption. Finally, the sensitivity function was multiplied by the observed spectrum for \CosmosI\ and 
renormalized for the different integration times between the standard star and the science. \CosmosII\ 
was flux-calibrated in the same manner with telluric standard stars  HIP41798 and HIP59351. 

The final flux calibrated spectrum was then constructed for each exposure by performing an unweighted summation of the spatial pixels of the IFU containing the 
object flux. We
then averaged each exposure's spectrum together. Noise estimates were determined by performing the 
identical procedure to five blank sky regions of the IFU. At every wavelength all noise spectra (five 
times as many as science spectra) were averaged together. The expected average was zero, and any 
non-zero component was also present in the science, thus this value was subtracted from the science flux. Note
that this final correction was negligible.
At every wavelength we determined the standard deviation of the noise. The uncertainty in our science 
flux was simply this standard deviation normalized by the square root of the number of science 
exposures. We stress that at no point were we subtracting science data from itself. The sky subtraction and noise 
were determined from multiple non-science regions of the IFU, and were thus statistically robust. 

Finally, any observed line at the expected systemic redshift was fitted with a Gaussian profile. To 
estimate line flux uncertainties, we repeated the fit for 6,000 synthetic spectra where each wavelength's 
flux was the sum of the flux observed and a random number from a normal distribution with standard 
deviation equal to the observed 1-$\sigma$ uncertainty in that wavelength's flux. Together these gave a 
distribution of possible line fluxes, where we took the mean value and the standard deviation to be the 
observed line flux and its 1-$\sigma$ uncertainty.
For lines that were not observed, 3-$\sigma$ upper limits on flux were found by following the method of 
\citet{Finkelstein11}, where a mock line profile was added to the spectrum at the expected wavelength 
and then repeatedly decreased in flux until the signal to noise ratio fell below the $3\sigma$ level.

\section{Final spectra}\label{reduction}
\begin{figure*}
\centering\leavevmode
\includegraphics[width={0.9\columnwidth}]{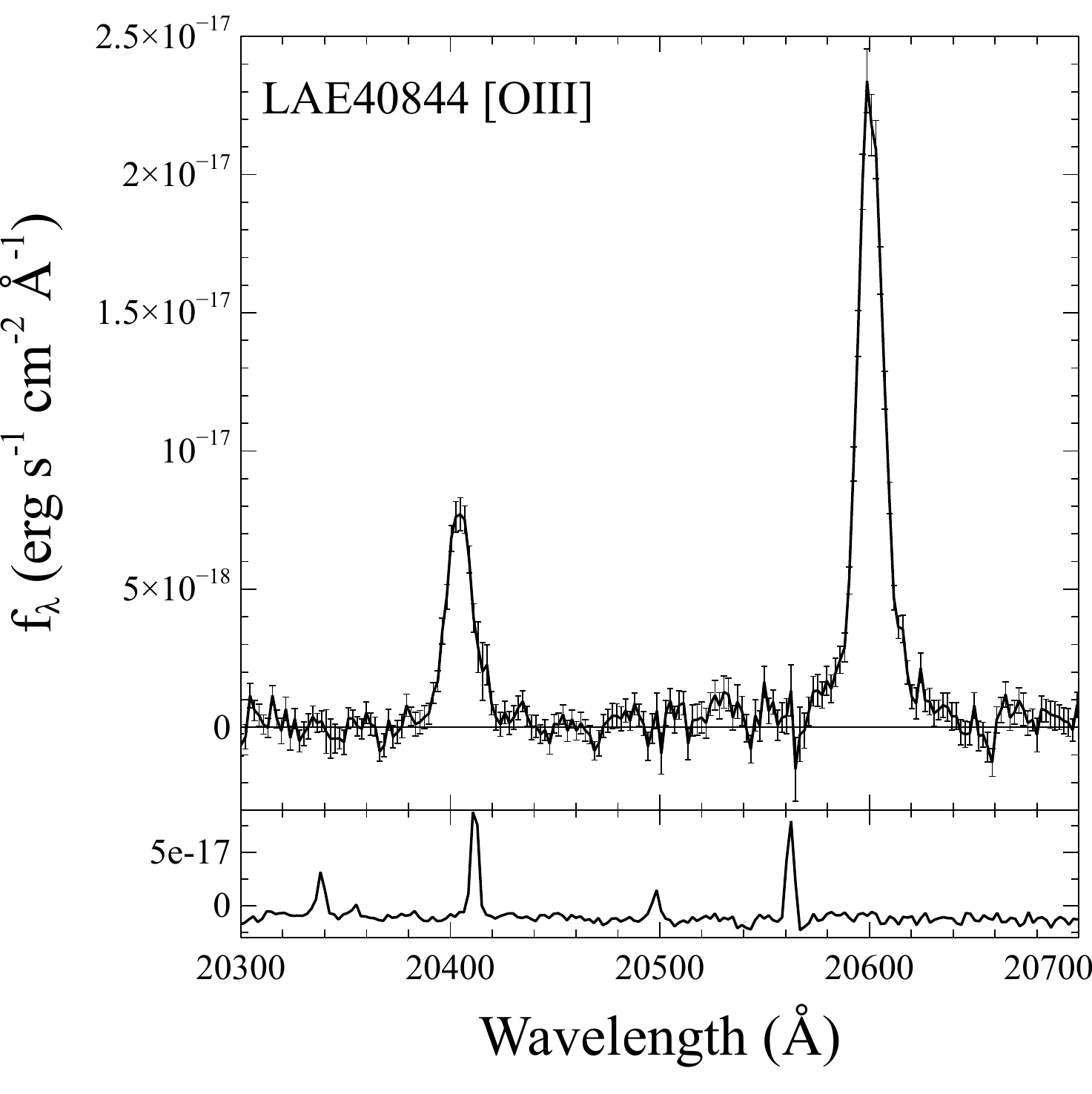} 
\includegraphics[width={0.9\columnwidth}]{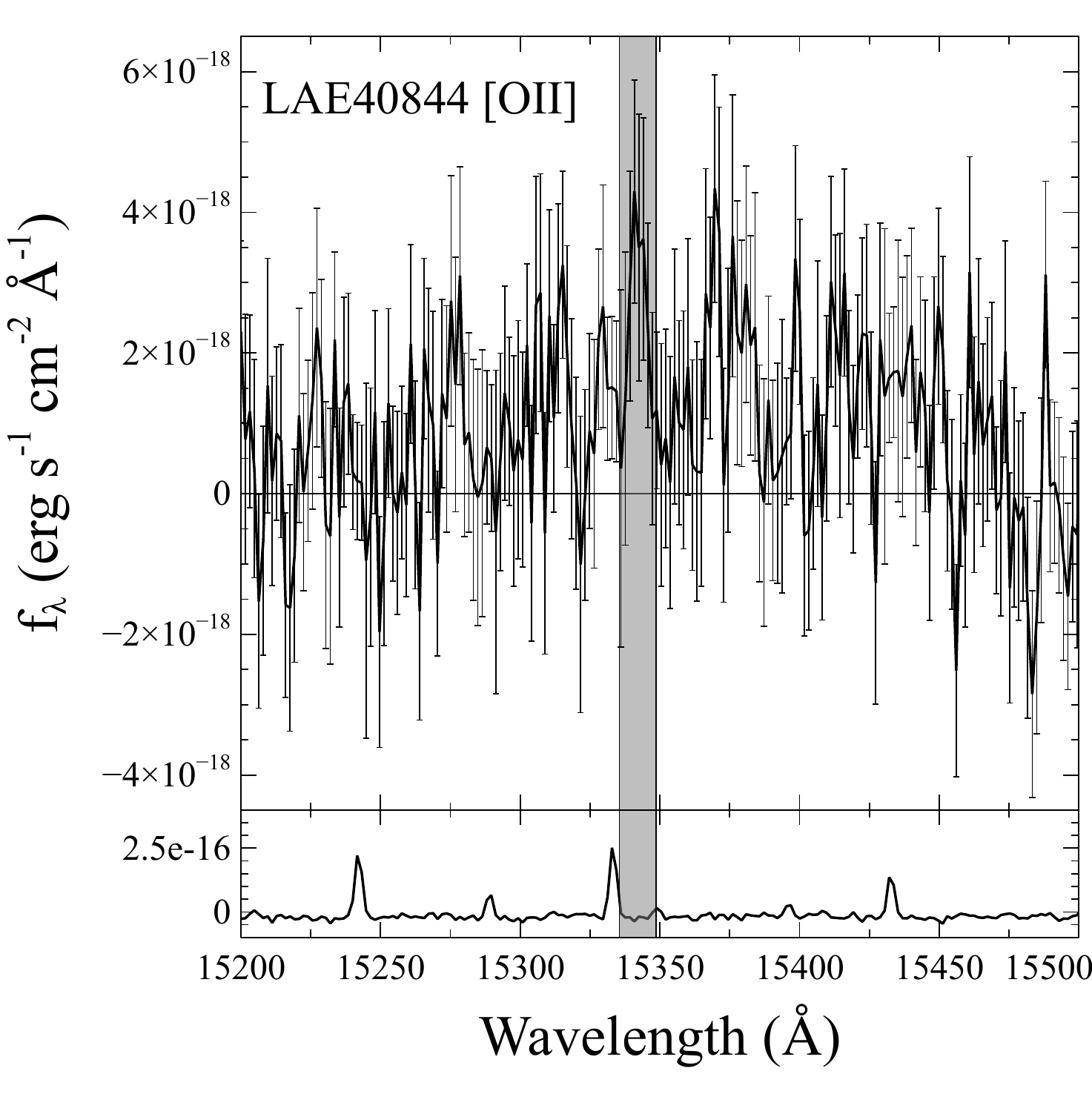} 
\includegraphics[width={0.9\columnwidth}]{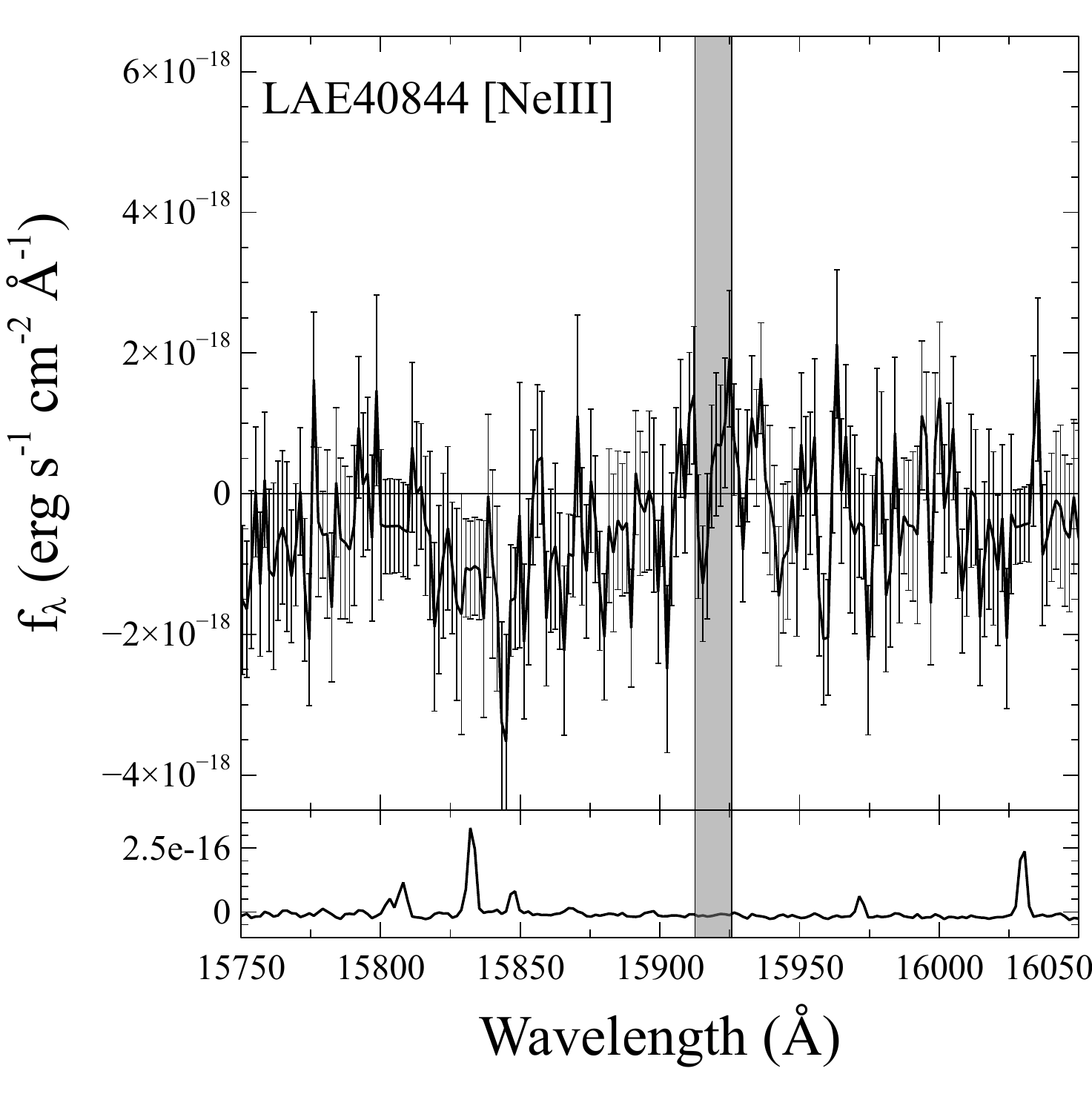}
\caption{\CosmosI\ spectra around [OIII] (top left), [OII] (top right), and [NeIII] (bottom). We include a sky 
spectrum below each plot. The [OIII] doublet is found at 20403.6 \AA\ and 20600.8 \AA\ yielding a 
redshift of $z_{[OIII]}=3.11330$. Thus we expect [OII] to be at 15342.1 \AA, shaded in grey with a width equal to the FWHM of the [OIII] line. We detect an
emission line at this wavelength with a signal-to-noise ratio of about 2. We expect [NeIII] to be at 15919.2 \AA,
shaded in grey  with a width equal to the FWHM of the [OIII] line, but have no detection. \vspace{1mm}}
\label{figCosmosI}
\end{figure*}

\begin{figure*}
\centering\leavevmode
\includegraphics[width={0.916\columnwidth}]{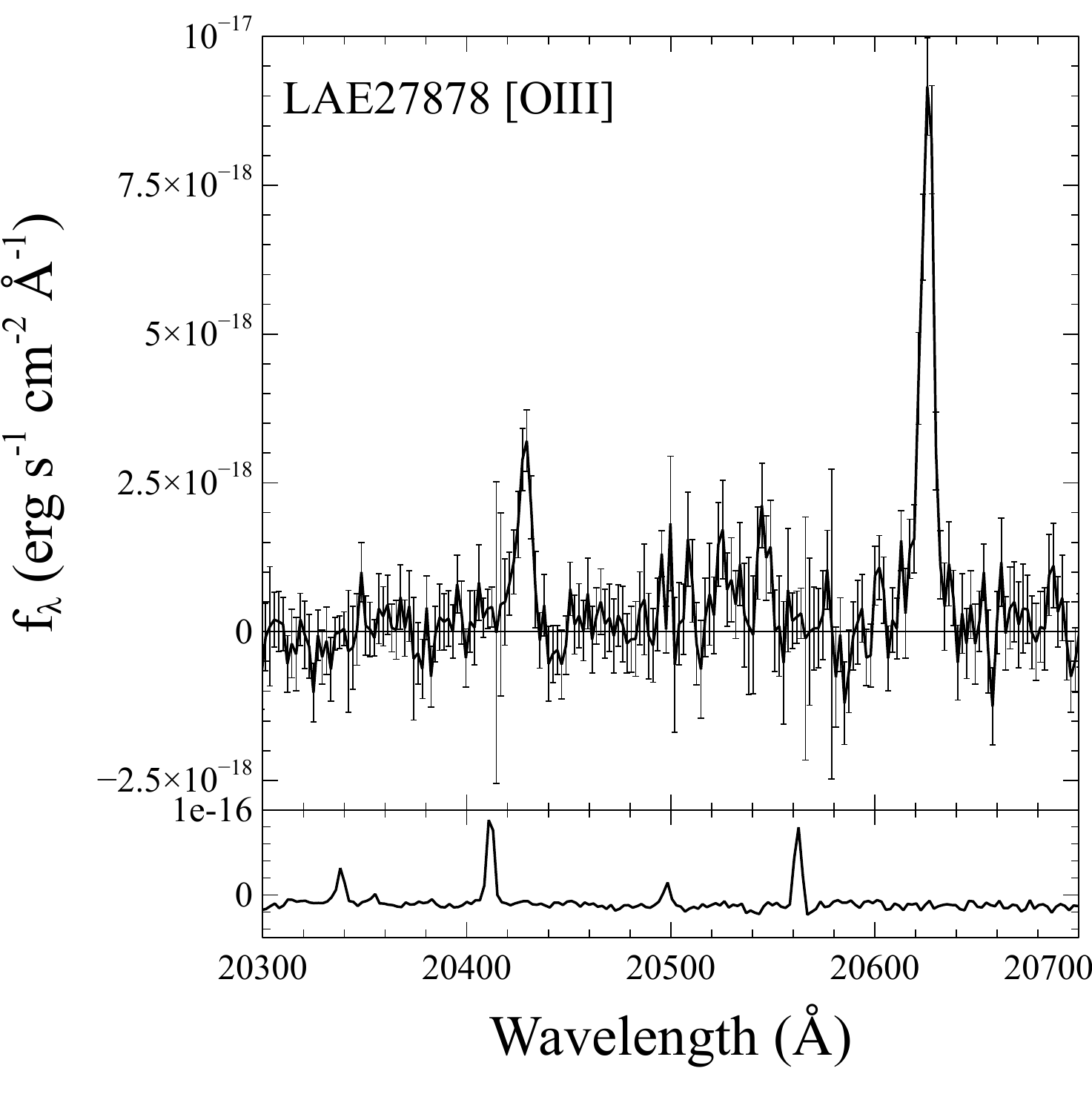} 
\includegraphics[width={0.91\columnwidth}]{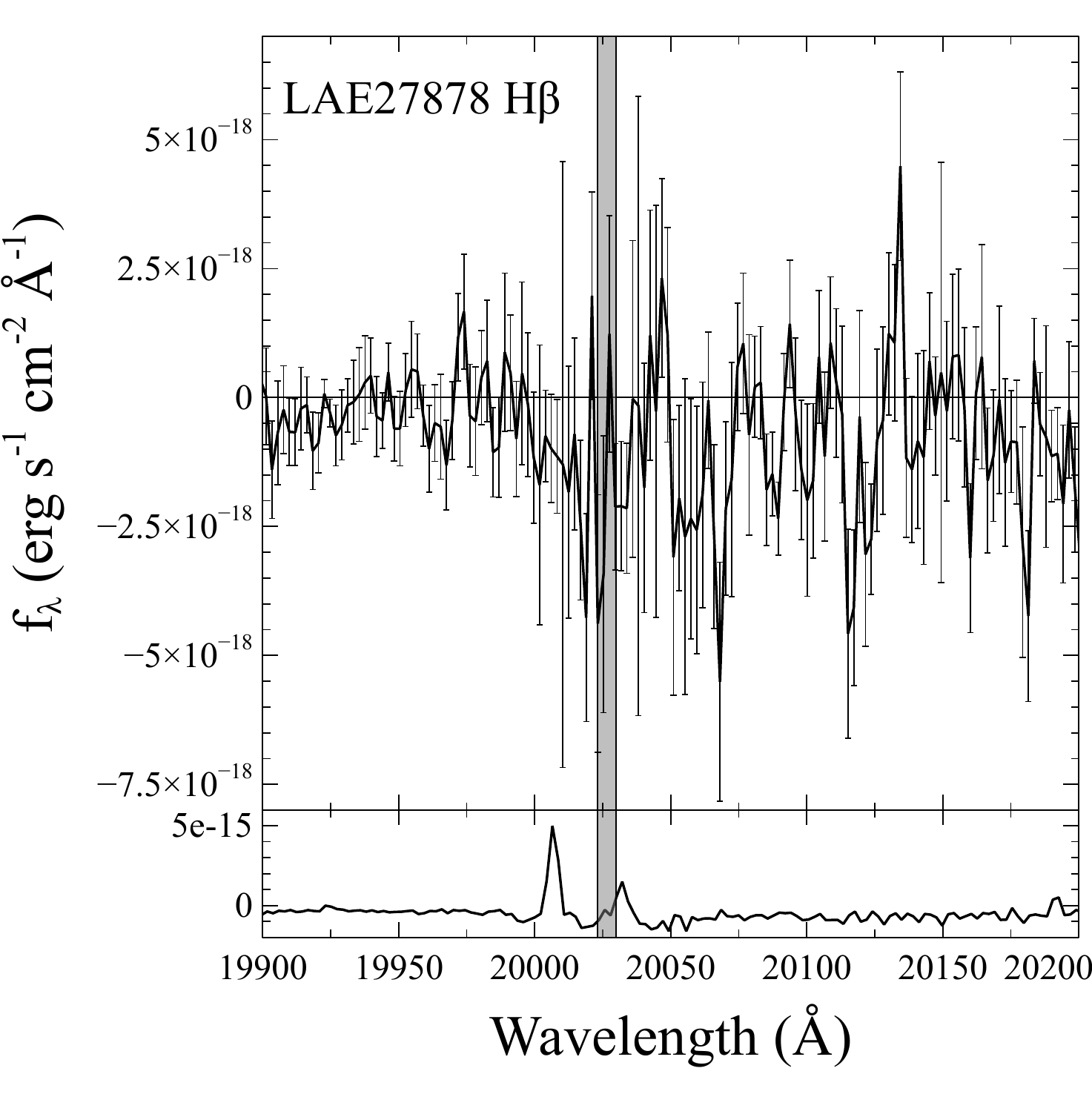}
\includegraphics[width={0.91\columnwidth}]{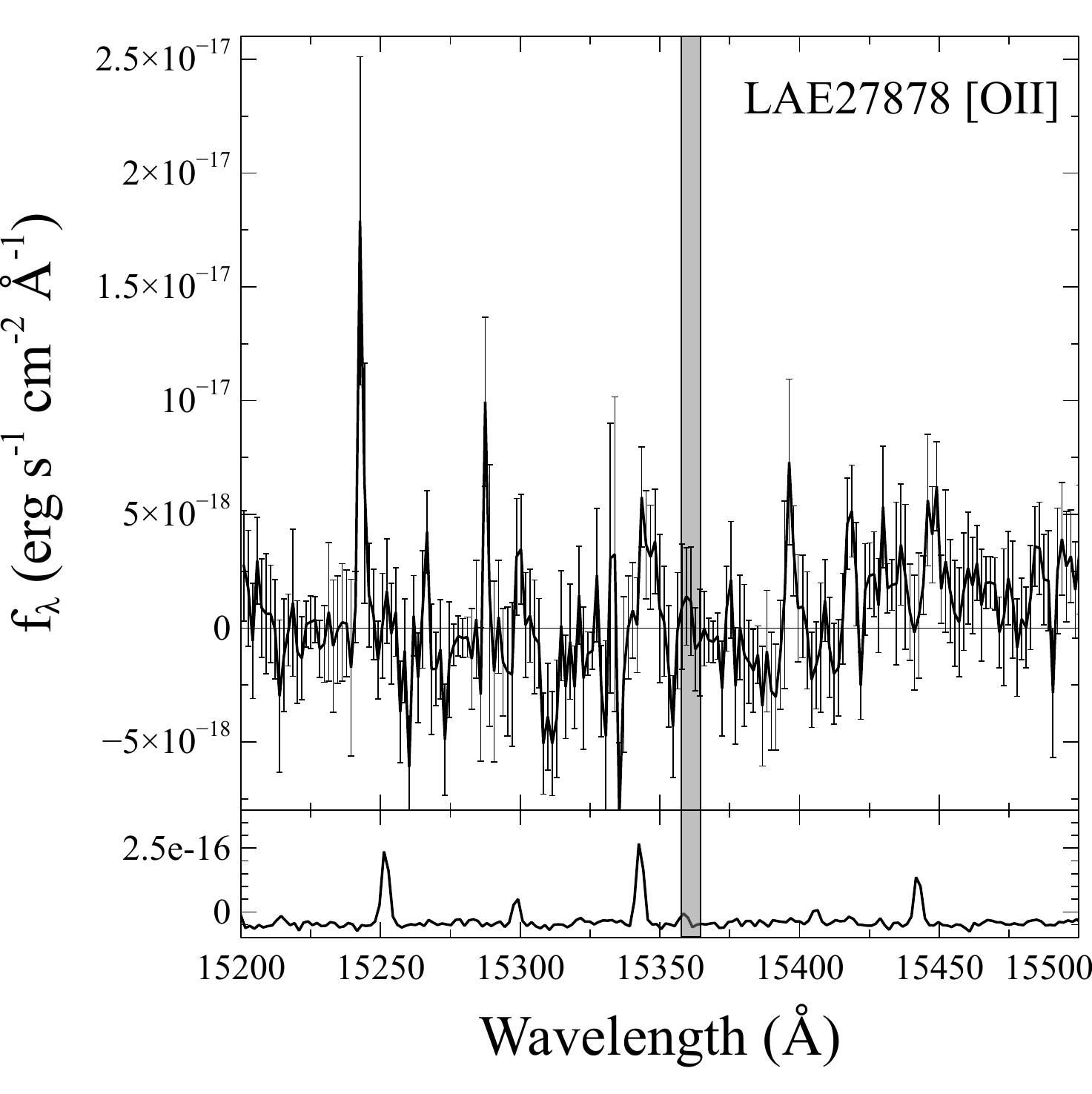} 
\includegraphics[width={0.91\columnwidth}]{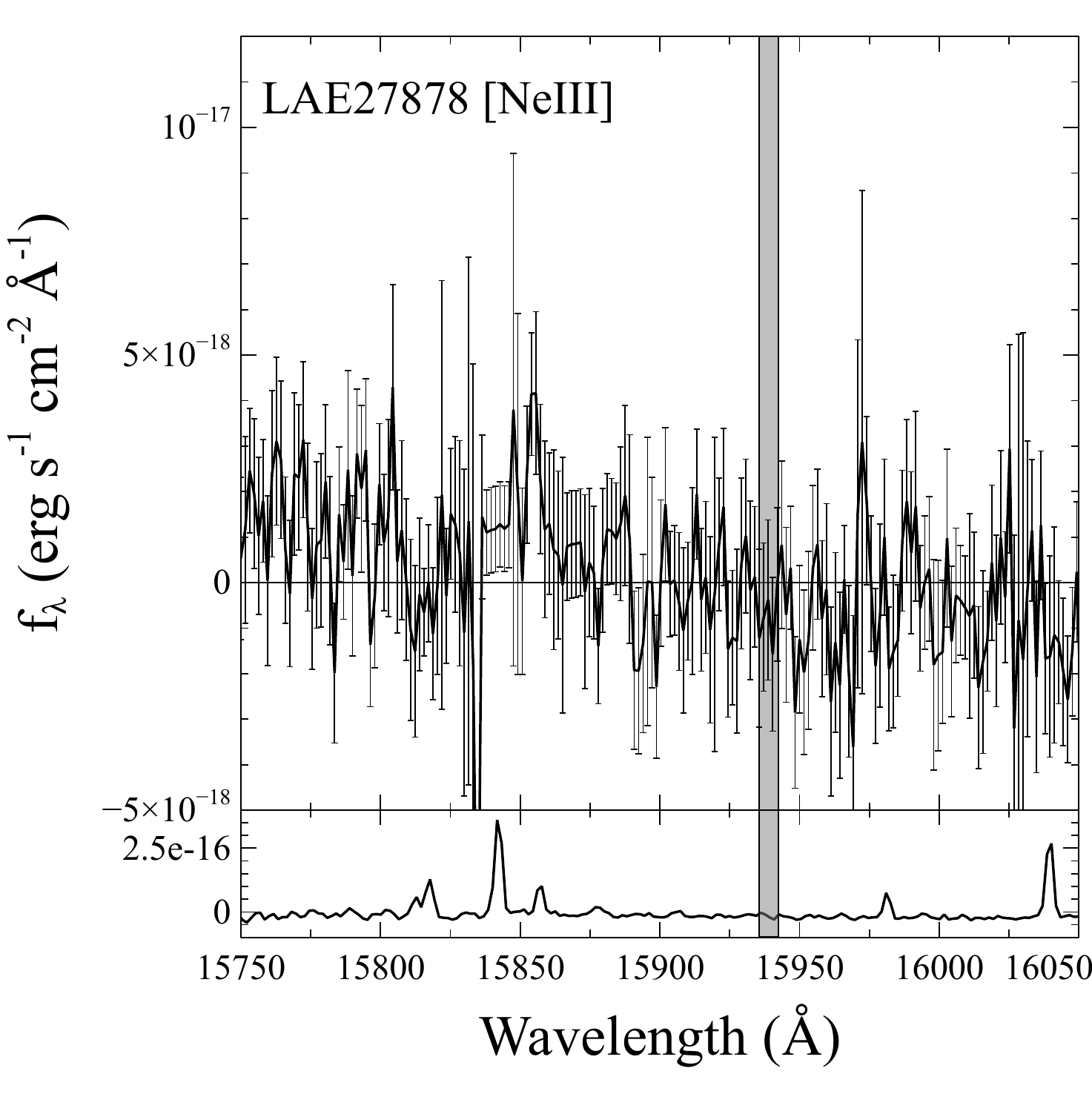}
\caption{\CosmosII\ spectra around [OIII] (top left), H$\beta$ (top right), [OII] (bottom left), and [NeIII] (bottom right). We include a sky spectrum below each plot. The [OIII] doublet is found at 20428.2 \AA\ and 
20625.7 \AA\ yielding a redshift of $z_{[OIII]}=3.11835$. Thus we expect H$\beta$ to be at 20026.5 
\AA, [NeIII] to be at 15938.9 \AA, and [OII] to be at 15361.2 \AA, all of which are shaded in grey  with a width equal to the FWHM of the [OIII] line. We have no detection and 
can only determine 3-$\sigma$ upper limits. \vspace{1mm}}
\label{figCosmosII}
\end{figure*}

We successfully detected [OIII] for both \CosmosI\ and \CosmosII\ with redshifts of $z_{[OIII]}=3.11330 \pm 
0.00011$ (see \fig{figCosmosI}) and $z_{[OIII]}=3.11835 \pm 0.00006$ (see \fig{figCosmosII}), 
respectively. We fit a Gaussian for the [OIII] 5007 \AA\ line, which had a reduced chi-squared value of 
1.4 and 1.1 for \CosmosI\ and \CosmosII, respectively, and determined the total line flux from this 
profile. The [OIII] 4959 \AA\ line was constrained to have the same redshift and line width as the 5007 \AA\ 
line. We then applied an extinction correction following \citet{Calzetti00} with an $E(B-V)$ value of 0.15 
and 0.10 for \CosmosI\ and \CosmosII, respectively, as determined by McLinden \etal\ in prep. (2013). 
These extinction values are the result of multi-band observations and SED fitting, and are consistent over small variations in
the best fit. The measured [OIII] flux from the 5007 (4959) \AA\ line was 32.33 $\pm$ 1.74 (11.90 $\pm$ 0.67)\Flux 
and 6.85 $\pm$ 1.06 (3.56 $\pm$ 0.51)\Flux for \CosmosI\ and \CosmosII, respectively.

\begin{deluxetable*}{lcc}
\tablecaption{Flux measurements and constraints for \CosmosI\ and \CosmosII.\label{tabred}}
\tablecolumns{3}
\tablewidth{0pc}
\tabletypesize{\footnotesize}
\tablehead{
	\colhead{Galaxy Characteristic} & 
	\colhead{\CosmosI} & 
	\colhead{\CosmosII} 
}
\startdata
$z_{Ly\alpha}$ & 3.11639 $\pm$ 0.00021 & 3.12051 $\pm$ 0.00021\\
Ly$\alpha$ line flux\tablenotemark{a} & $36.1^{+2.35}_{-2.47}$ & $9.41^{+ 1.42}_{-1.63}$\\
$z_{[OIII]}$ & 3.11330 $\pm$ 0.00011 & 3.11835 $\pm$ 0.00006\\
$[$OIII$]$ line flux ($\lambda = 5008.240$ \AA)\tablenotemark{a} & 32.33 $\pm$ 1.74 & 6.85 $\pm$ 1.06\\
$[$OIII$]$ line flux ($\lambda = 4960.295$ \AA)\tablenotemark{a} & 11.90 $\pm$ 0.67 & 3.56 $\pm$ 0.51\\
$[$OII$]$ line flux\tablenotemark{a} & 3.05 $\pm$ 1.63 & $<$ 2.9 \\
$[$NeIII$]$ line flux\tablenotemark{a} & $<$ 1.8 & $<$ 2.6\\
Ionization Parameter, $q$\tablenotemark{b} & $> 2 \times 10^8$ & $ >4 \times 10^7$\\ 
Metallicity, Z\tablenotemark{c} & $< 1$ & -\\
H$\beta$ line flux\tablenotemark{a,d} & $>3.4 $ & $> 0.6$\\  
$f_{esc,Ly\alpha}$ & $< 46$\% & $<74$\%\\ 
\enddata
\tablenotetext{a}{$10^{-17}\mbox{ erg g s}^{-1}\mbox{ cm}^{-2}$}
\tablenotetext{b}{$\mbox{cm s}^{-1}$ - assuming a minimum metallicity of $Z=0.05\Zsun$, see 
\sect{metal} \vspace{1mm}}
\tablenotetext{c}{In units of Z$_\odot$}
\tablenotetext{d}{Constraints inferred from the observed [OIII] flux and the upper limit on [OIII]/H$\beta$ from our models.}
\end{deluxetable*}

Following the same technique as for the [OIII] 4959 \AA\ line, we detected [OII] at about the 2-$\sigma$ level for \CosmosI\ (see grey region in 
\fig{figCosmosI}). After applying the extinction correction we measured the [OII] flux to be 3.05 $\pm$ 1.63\Flux.
We did not detect [OII] from \CosmosII, thus, following the procedure outlined in \sect{sample}, we determined 3-$\sigma$ upper limits on the [OII] line flux to be 2.9\Flux for 
\CosmosII. 

We did not detect [NeIII] 3869 \AA\ in either \CosmosI\ or \CosmosII. We thus determined
a 3-$\sigma$ upper limit on the [NeIII] line of 1.8 and 2.6\Flux\ for \CosmosI\ and \CosmosII, respectively.
We only had H$\beta$ coverage for \CosmosII\ (see middle of 
\fig{figCosmosII}). Unfortunately, we were unable to completely compensate for the strong telluric 
variations at this wavelength and had no detection. We got a 3-$\sigma$ upper limit on the H$\beta$ 
flux of 4.3\Flux. We recognize that time variation in the telluric absorption may be unaccounted for in our sky subtraction 
technique, thus we limit any further conclusions based on this upper limit. A complete summary of the measured and determined line fluxes for \CosmosI\ and 
\CosmosII\ can be seen in \tabl{tabred}.

The single detection of [OIII] and constraint of [OII] for \CosmosII\ allows us to make few conclusions for this object. The 
lower limit for this ratio implies an ionization parameter consistent with the models of \citet{Levesque10}, although a larger
ionization parameter is possible. We explore this more fully in \sect{results}. The detections of both [OIII] and [OII] for
\CosmosI\ demonstrate that this galaxy has an ionization parameter larger than those considered in \citet{Levesque10}. To better understand this galaxy, and other high ionization galaxies 
discovered recently, we present models in this regime, continuing the work of \citet{Levesque10}. \vspace{5mm}



\section{Ionization Models}\label{models}
We created models to compare with observed emission lines from galaxies, constraining their
gas-phase metallicities and ionization parameters. Previous works (\emph{e.g.} Kewley \etal\ 2002; Nagao \etal\ 2006; Levesque \etal\ 2010) have presented models with ionization parameters in the range of $5\times10^6$ - $4\times10^8\cms$. However, larger ionization parameters, although uncommon in the local universe (e.g., Rigby \& Rieke 2004; Lilly \etal\ 2003), may very well be present in the early universe around stronger, younger starbursts. Indeed, recent observations (e.g., Fosbury \etal\ 2003; Richard \etal\ 2011; Erb \etal\ 2010; Xia \etal\ 2012; Jaskot \& Oey 2013; Nakajima \etal\ 2013) have demonstrated that these earlier galaxies undergoing significant starburst do have larger ionization parameters. This work extends the existing models of \citet{Levesque10} to higher values of ionization parameter.

To simulate photoionization regions we took the Starburst99 (Leitherer et al. 1999; V\'{a}zquez \& 
Leitherer 2005) outputs of \citet{Levesque10} and used these as input for the photoionization code 
Mappings III (Binette et al. 1985; Sutherland \& Dopita 1993). We wished to investigate models with 
larger ionization parameters, extending earlier work to best constrain the physical parameters of \CosmosI. 
All Starburst99 models were previously described in \citet{Levesque10}.

The Mappings code was developed by \citet{Binnette85} to model shocks and photoionization regions. 
\citet{Sutherland93} improved upon this code, which was used in \citet{Dopita00} and
\citet{Kewley01}. These were followed by more improvements, including a more sophisticated treatment of dust 
\citep{Groves04}, the effects of absorption, charging, and 
photoelectric heating by the grains. These improvements are explained in more detail in 
\citet{Groves04} and \citet{Snijders07} and were used in \citet{Kewley06} and \citet{Levesque10}. 
Given the artificial ionizing FUV spectrum from Starburst99 we adopted a plane-parallel nebular 
geometry with the same range of densities as \citet{Levesque10} but with a range of larger ionization 
parameters. Using these parameters in Mappings III, we computed a grid of plane-parallel isobaric 
photoionization models.

We studied both continuous and instantaneous star formation histories with model ages of 6 Myr and  
0, 1, and 3 Myr, respectively. We considered metallicities of 0.05, 0.2, 0.4 
and 1 solar (using the solar value of $\log(O/H)+12=8.90$ presented in McGaugh 1991), and we adopted
the HIGH mass-loss tracks from the Geneva models. Given these outputs 
we then generated model line fluxes using Mappings III models with hydrogen densities of 10 and 100 
$\cms$ and ionization parameters of 
$4$, 
$6$, 
$9$, 
$15$, 
$30$, 
$45$, 
$60$, 
$90$, 
$150$, and 
$300 \times 10^{8}\cms$. Recall that the ionization parameter is the ratio between 
the ionizing photon flux and the hydrogen density, and is a proxy for the fastest speed an ionizing front 
can attain. The largest ionization parameter was chosen to agree with the work of \citet{Groves04} where $q_{\rm max} = \emph{c}$, 
however we note that the samples of \citet{Rigby04} and \citet{Lilly03} illustrate that galaxies, in general, are not seen 
with $q$ values above $2.0\times 10^{8}\cms$. These much larger ionization 
parameters are expected for more metal-poor galaxies in their first cycles of star-formation.

In the following subsections we briefly discuss the line diagnostics line ratios appropriate for ground-based observations of $z\sim3$ galaxies. This limits us to the 
[OIII]$\lambda 5007$/[OII] $\lambda 3727$, [OIII]$\lambda 5007$/H$\beta$, and $R_{\rm 23}$ line ratios, and we present them in various
combinations. We also discuss in Appendix A the  [NeIII]$\lambda 3869$/[OII]$\lambda3727$, [NII]$\lambda 6584$/H$\alpha$,  and [NII]$\lambda 6584$/[OII]$\lambda 3727$ line ratios, presenting them in various combinations, and discuss these
new models' implication for other objects in the literature.

\subsection{[OIII]$\lambda 5007$/[OII] $\lambda 3727$}
The [OIII]$\lambda 5007$/[OII] $\lambda 3727$ line ratio is described in detail in \citet{Levesque10}. This ratio is 
commonly used as an ionization parameter diagnostic, although it is weakly dependent on metallicity \citep{Kewley02},
with larger values corresponding with larger ionization parameter, or lower abundances (see for example Figures \ref{figOIII_OII_vs_NII_OII} and \ref{figOIII_OII_vs_NII_Ha}).

\subsection{[OIII]$\lambda 5007$/H$\beta$}
The [OIII]$\lambda 5007$/H$\beta$ line ratio is well described for the case of low to medium ionization 
parameter in \citet{Levesque10}. In this case, the inferred abundance is particularly sensitive to the ionization parameter 
for ``normal'' ionization parameters and sub-solar abundances.  For ionization parameters $q \ga 5\times 10^9$ cm s$^{-1}$, 
the [OIII]/H$\beta$ ratio saturates, making it a more useful diagnostic of sub-solar abundances
(see for example Figures \ref{figOIII_Hb_vs_NeIII_OII} and \ref{figOIII_Hb_vs_NII_Ha}). At both large and small ionization parameters this ratio is double-valued
with respect to abundance. Of special note, these lines are very similar in wavelength, thus the ratio will not suffer from 
significant reddening effects.

\subsection{$R_{23}$}
The $R_{23}$ diagnostic, defined as log$_{\rm 10}$$\left( ([OII]\lambda 3727+ [OIII]\lambda 4959,5007)/H\beta\right)$ \citep{Pagel79}, is described 
in detail in \citet{Kewley02}. It is commonly used as a metallicity diagnostic, although it is double-valued and dependent on the ionization 
parameter, thus it must be combined with another diagnostic to determine the metallicity and ionization parameter correctly (see for example 
\fig{figOIII_OII_vs_R23}). It is particularly useful at higher redshifts, where H$\alpha$, [NII] $\lambda 6584$, and [SII] $\lambda 6717,31$ are too far redshifted to be easily observed.
 
\subsection{Diagnostic Diagrams}
The motive for these models was to better constrain the physical parameters of
high-ionization parameter galaxies. To best accomplish this, typically two
line diagnostics are compared via emission line diagnostic diagrams.

\begin{figure*}[t]
\centering\leavevmode
\includegraphics[width={1.96\columnwidth}]{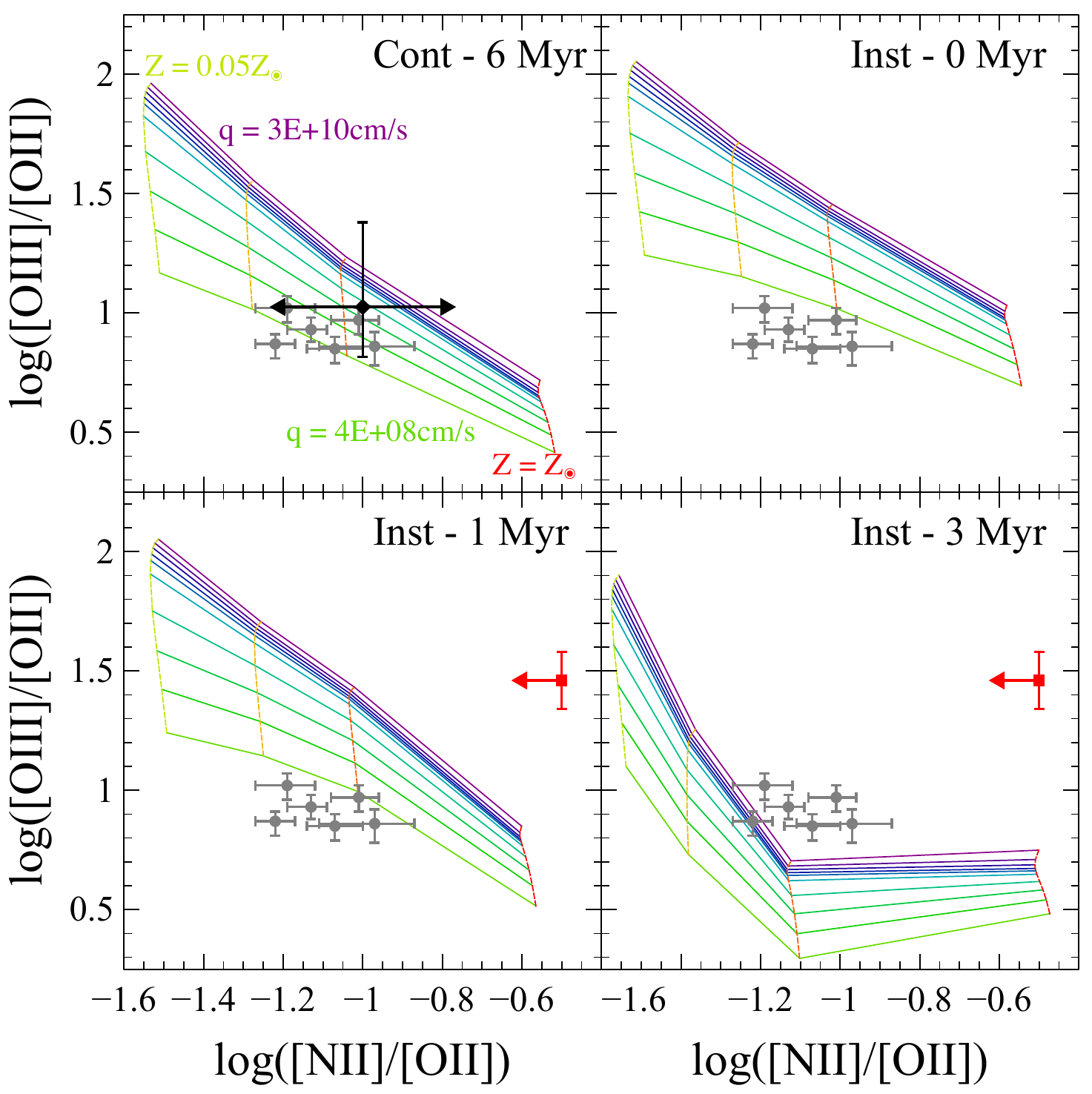} 
\caption{\footnotesize{Presented here are the [OIII]$\lambda 5007$/[OII] $\lambda 3727$ values versus the [NII] 
$\lambda 6594$/[OII] $\lambda 3727$ (discussed more in Appendix A) values from the photoionization models of continuous star-formation for a 6 
Myr population (top left), and the photoionization models of instantaneous star-formation 
for a 0 Myr (top right), 1 Myr (bottom left) and 3 Myr (bottom right) population. A stellar population with a continuous star-formation history reaches 
equilibrium around 5 Myr, and then is constant. The near-vertical dashed lines are of constant metallicity, 
with the top left being 0.05 solar, followed by 0.2, then 0.4, and the bottom right being solar. The solid lines are of constant 
ionization parameter, with $4\times 10^8\cms$ for the lower line, followed by 
$6$, 
$9$, 
$15$, 
$30$, 
$45$, 
$60$, 
$90$, 
$150$, and 
$300 \times 10^{8}\cms$ for the higher line. Consistently we see a transition present in the 3 Myr instantaneous 
models where the most abundant cases have sufficiently cooled, limiting the impact of abundance on the [OIII]/[OII] diagnostic. 
We include the [OIII]/[OII] constraints determined for \CosmosI\ (black diamond), which require Z $<\Zsun$. Note that we have no constraint for 
[NII]/[OII]. Also included are observations and 1-$\sigma$ error bars of CDFS-3865 by \citet{Nakajima13} (red square) and the "Green Pea" galaxies 
from \citet{Jaskot13} (gray circles),
allowing for a range of ionization parameters and sub-solar metallicities.
\vspace{3mm}}}
\label{figOIII_OII_vs_NII_OII}
\end{figure*}

\begin{figure*}[t]
\centering\leavevmode
\includegraphics[width={1.96\columnwidth}]{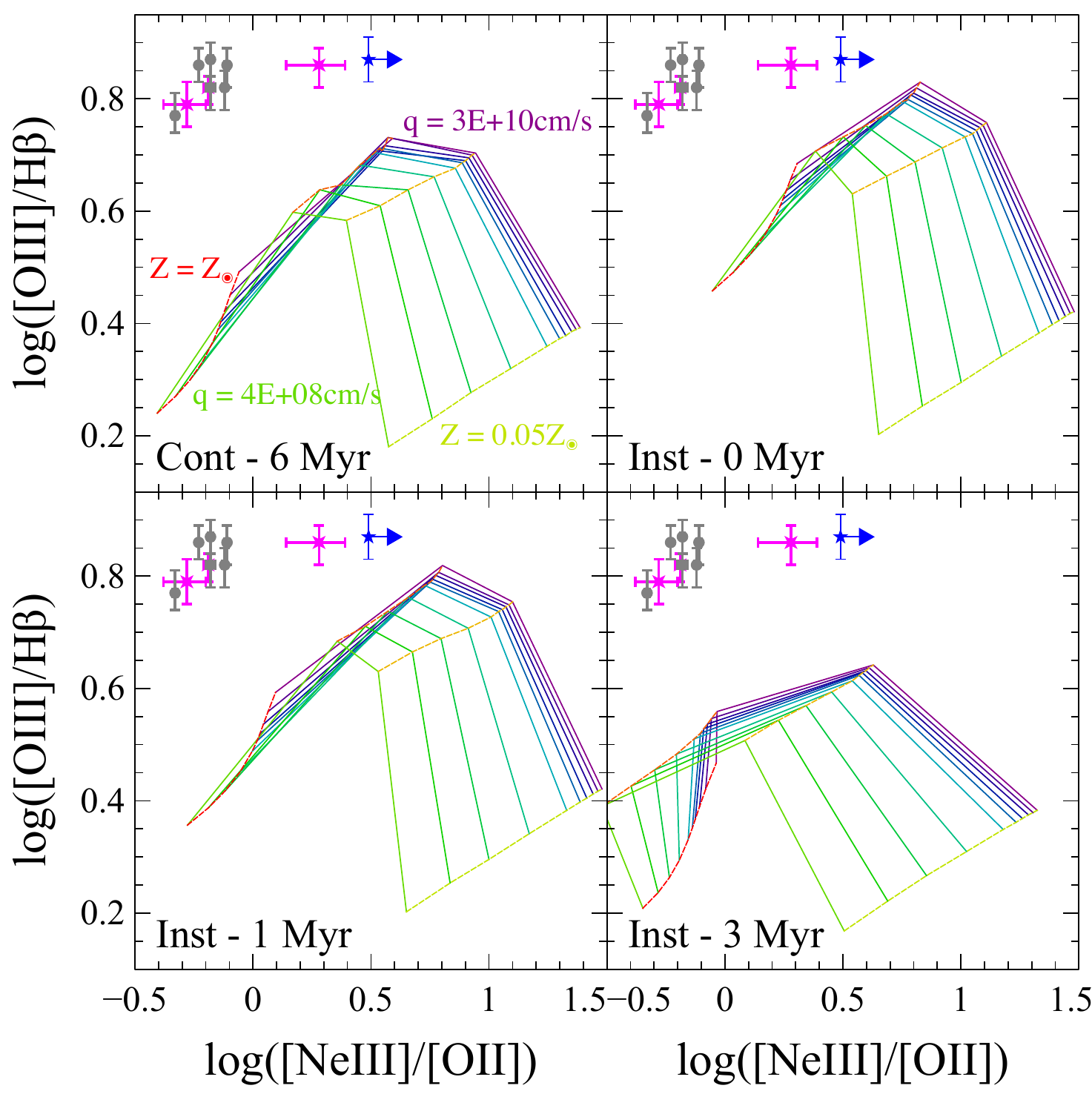} 
\caption{\footnotesize{Similar to \fig{figOIII_OII_vs_NII_OII}, we present here the [OIII]$\lambda 5007$/H$\beta$ values versus the [NeIII] 
$\lambda 3869$/[OII] $\lambda 3727$ (discussed more in Appendix A) values. Both line diagnostics in this plot are 
insensitive to reddening. Included are constraints and 1-$\sigma$ error bars 
of the Lynx arc by \citet{Fosbury03} (blue star), objects 20201, 31362, and 823LZ from Xia \etal\ in prep. 2013 (magenta pointed cross), and \citet{Jaskot13} (gray circles), requiring 
possibly very large ionization parameters and a range of metallicities. The offset observed between these objects and model grids suggests that additional physics may need
to be included in the modeling. We explore this further in \sect{results}. \vspace{1mm}}}
\label{figOIII_Hb_vs_NeIII_OII}
\end{figure*}

\begin{figure*}[t]
\centering\leavevmode
\includegraphics[width={1.96\columnwidth}]{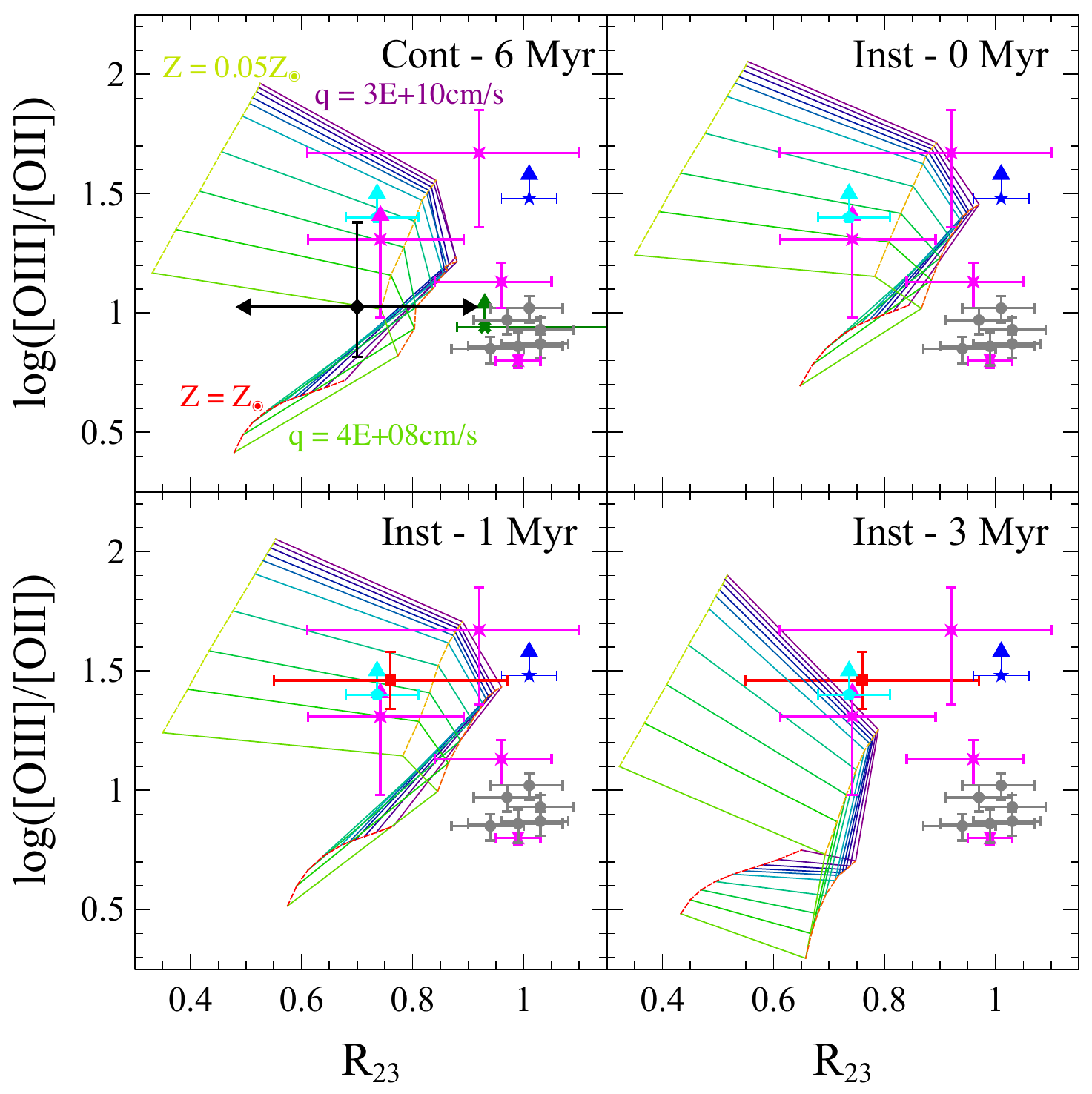} 
\caption{\footnotesize{Similar to \fig{figOIII_OII_vs_NII_OII}, we present here the [OIII]$\lambda 5007$/[OII] $\lambda 3727$ values versus the $R_{23}$ 
diagnostic, $\log_{10} \left( ([OII]\lambda 3727+ [OIII]\lambda 4959,5007)/H\beta\right)$. 
Included are the constraints and 1-$\sigma$ error bars of \CosmosI\ from this work (black diamond), the Lynx arc by \citet{Fosbury03} (blue star),  
BX418 from \citet{Erb10} (green flat cross), Sextet by \citet{Richard11} (cyan pentagon), object 246 from \citet{Xia12} and 20201, 31362, and 823LZ from Xia \etal\ in prep. 2013 (magenta pointed cross),
CDFS-3865 by \citet{Nakajima13} (red square), and the extreme green peas in \citet{Jaskot13} (gray circles). Note that
\CosmosI\ has no constraint on R$_{23}$ since we are unable to detect H$\beta$. Again we have many observations that are not consistent with these modes. We explore this further in \sect{results}. \vspace{1mm}}}
\label{figOIII_OII_vs_R23}
\end{figure*}

Figures \ref{figOIII_OII_vs_NII_OII}-\ref{figOIII_OII_vs_R23} combine in multiple pairings the
different line ratios discussed above. We select these pairings to highlight the line combinations observable with ground-based spectroscopy of $z\simeq3$ LAE, 
and in particular the two LAE that we observe.
Other pairings that may prove useful with future observations can be found in Appendix A. Plotted are the models with a density of 10 cm$^{-3}$.  
In all cases the models with a density of 100 cm$^{-3}$ are very similar to the 10 cm$^{-3}$ models, 
and are omitted here for brevity\footnote{All models can be found at http://www.public.asu.edu/$\sim$mlricha4/HII\_Models.tgz}. 
Dashed lines of constant metallicity and solid lines of constant ionization parameter are plotted. 

Included in these figures are observations of 
high ionization parameter galaxies, including the observations of \CosmosI\ from this work. Where possible, we plot each object on the panel whose model 
star-formation history best matches that reported for the object in the literature.  In some cases, published information on age and star formation history is 
not sufficiently detailed for us to do this (notably, objects from Richard et al. 2011, Xia et al. 2012, and 2013 in prep, and Jaskot \& Oey 2013).  In these 
cases, we plot the objects on all panels. It is important to note that there is significant variation of the ionization parameter and metallicity grids from one 
stellar age to the next. Thus it is essential to know the age \emph{a priori} to determine how these models, combined with observations, constrain the 
physical parameters of a galaxy. Without the age constrained, it would require several emission line diagnostics to determine
a self-consistent solution to age, metallicity and ionization parameter.

\section{Results \& Discussion} \label{results}
In this section we discuss how our photoionization models can be used to infer constraints on the 
metallicity and ionization parameter for \CosmosI, \CosmosII, and what further inferences can be made. How our models describe other high-ionization parameter galaxies is discussed in Appendix A.
There are a few notable examples where specific combinations of line diagnostics yield model grids that are inconsistent with observations.
All combinations that include [OIII]/H$\beta$ or R$_{23}$ have the observations slightly offset from the bend in the double-valued diagnostic
(e.g., Figures \ref{figOIII_Hb_vs_NeIII_OII}, \ref{figOIII_OII_vs_R23}, \ref{figOIII_Hb_vs_NII_Ha}). We see similar offsets in the diagrams with [NeIII]/[OII].
Such inconsistencies were also observed by
\citet{Levesque10} when combining [OIII]/H$\beta$ with [SII]/H$\alpha$. Earlier models, such as \citet{Kewley01}, were even less able to
predict such large line diagnostics, and predicted that including harder far-ultraviolet (FUV) photons from a more accurate modeling 
of stellar atmospheres would account for this discrepancy. Specifically, the deficiency in high-energy photons results in underpredicting
these species' line fluxes. Although \citet{Levesque10} showed that their models, including the new atmospheric
models of \citet{Hillier98} and \citet{Pauldrach01} in Starburst99, were better able to predict large emission line diagnostic, there were still
some shortcomings in the models. These same shortcomings are apparent in our work. An even better treatment of the FUV is necessary to
amend these model grids. Further, our models do not take into account the rotation of stars. After a few Myr, stellar rotation has been shown to produce
a harder ionizing spectrum at solar metallicities \citep{Levesque12}, and this effect should be even more pronounced at lower metallicities. It is therefore
not surprising that such an offset would be visible for line diagnostics associated with [OIII] and [NeIII], both of which span a broad range of ionizing wavelengths.
While this does not diminish their efficacy as diagnostics, these offsets do effectively illustrate the need for additional improvements to the models used in 
developing such diagnostics.

\subsection{\CosmosI}\label{metal}
Given our inability to observe \Hb\ for \CosmosI, we would naively think, based on earlier models of low ionization parameter galaxies, that we'd be unable to
constrain either the metallicity or the ionization parameter. However, similar to the procedure in \citet{Erb10}, a comparison with the emission-line diagnostics of \citet{Levesque10} reveals that for our sufficiently large [OIII]/[OII] ratio, \CosmosI\ must be either very low metallicity, or at higher ionization than they consider. From SED fitting in McLinden et al. in prep (2013) we
know that \CosmosI\ has been undergoing near constant star formation for the last 7 million years, thus we will compare it with our continuous star formation models. Our new models reveal that at solar metallicity, 
even our largest ionization parameter models would not predict such a large [OIII]/[OII] ratio (\emph{e.g.} Figures \ref{figOIII_OII_vs_NII_OII}, \ref{figOIII_OII_vs_R23}, \ref{figOIII_OII_vs_NII_Ha}). Even larger 
ionization parameters are unlikely to make a difference as its effect asymptotes near $q\sim c$. Note these figures give no constraint on the 
horizontal axis. We stress that this upper limit on oxygen abundance is entirely dependent on the knowledge of the star
formation history of \CosmosI, as a zero-age population from an instantaneous burst of star formation does allow for such a large [OIII]/[OII] ratio at solar abundances, while a 3 Myr population from an instantaneous burst of star formation would require even stricter upper limits on the abundance.

Assuming a metallicity above $0.05\Zsun$, combining our models with those of \citet{Levesque10} requires \CosmosI\ to have an ionization 
parameter of at least  $2\times 10^8\mbox{ cm s}^{-1}$ to explain the observed line ratio of [OIII]/[OII]. Allowing for 
an even more metal-poor environment would likely allow for a lower ionization parameter.
\citet{Rigby04} compared multiple studies of the ionization parameter, using an average 
value for starburst galaxies of $10^{-2.3}\mbox{ }c = 1.5 \times 10^8 \mbox{ cm s}^{-1}$, and an 
upper limit of about $10^{-0.7}\mbox{ }c = 6 \times 10^9 \mbox{ cm s}^{-1}$. \citet{Lilly03} observed 66
$0.47<z<0.92$ UV-luminous galaxies and found the largest log([OIII]/[OII]) ratio to be roughly 0.6.
 We thus conclude the \CosmosI\ has a larger ionization parameter than the average local starburst galaxy 
\citep{Rigby04} and the $z<1$ high-UV galaxies of \citet{Lilly03}. From \fig{figOIII_OII_vs_NII_OII}, we conclude that 
\CosmosI\ has a metallicity less than solar, and likely less than $\sim 0.6$ solar, if we assume the same maximum ionization parameter of \citep{Rigby04}. Comparisons
with \citet{Maiolino08} who fit the observed line diagnostics of the low-metallicity galaxies of \citet{Nagao06} and the SDSS DR4 galaxies with their
inferred abundances show that \CosmosI\ is expected to have a low metallicity, with $Z=0.04^{+0.09}_{-0.04}\Zsun$, accounting for a spread of 0.2-0.3 dex.

\begin{figure*}[t]
\centering\leavevmode
\includegraphics[width={1.96\columnwidth}]{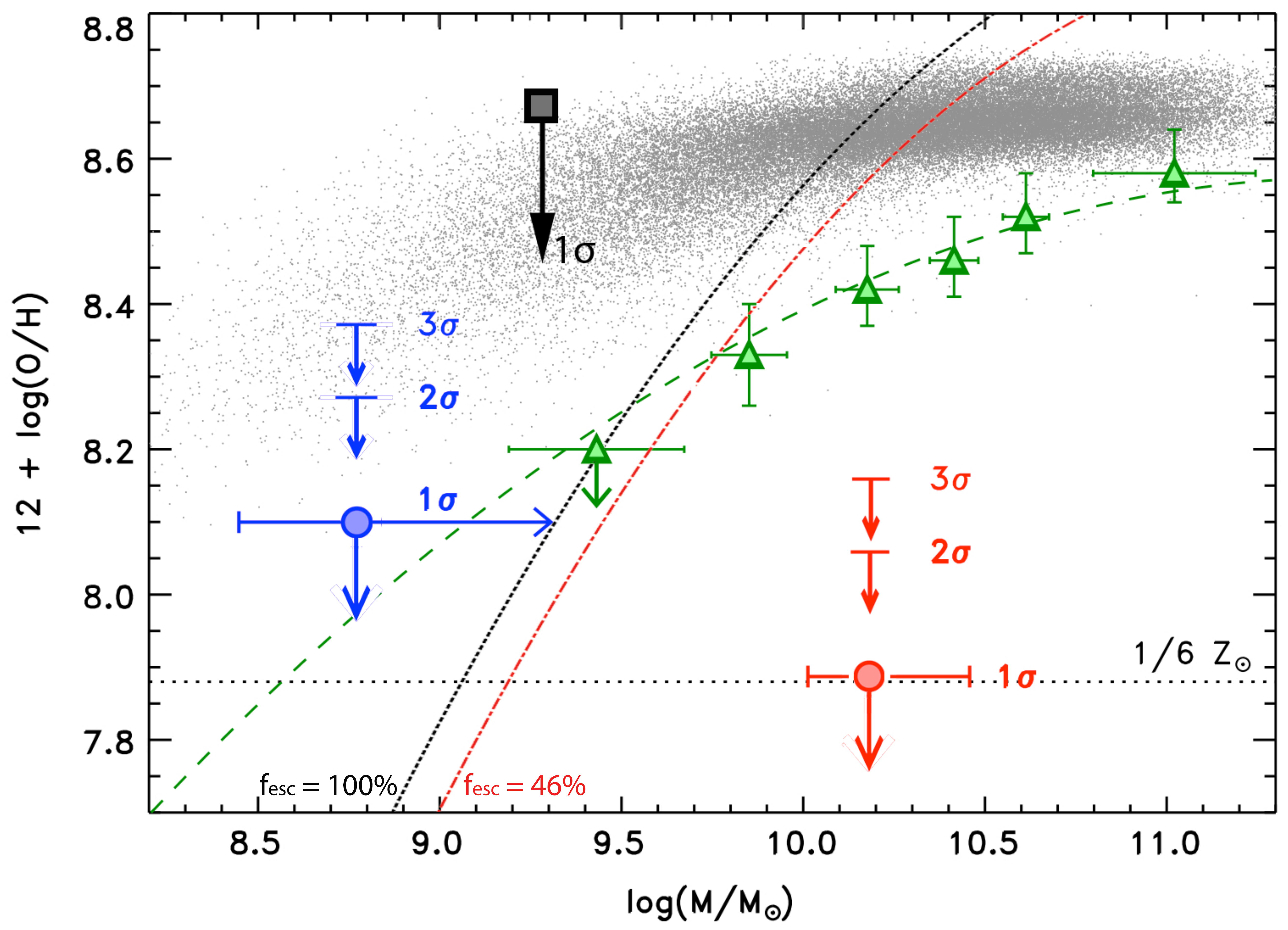} 
\caption{Figure taken from \citet{Finkelstein11}, with \CosmosI\ (black square) plotted with its oxygen 
abundance, determined in this work, vs. stellar mass, determined in McLinden \etal\ in prep. (2013). Also plotted are the $z\sim 0.1$ SDSS sample of \citet{Tremonti04} (gray 
points), the $z\sim 2.3$ star-forming galaxies in \citet{Erb06} (green triangles with a fit given by a green 
dashed line), and the two LAEs (z=2.29 and 2.49) of \citet{Finkelstein11} whose metallicities were 
constrained (red and blue circle). \CosmosI\ is in agreement with the $z\sim2.3$ sample of 
\citet{Erb06}. Note that we have included two
lines of constant star-formation taken from the FMR for $z\sim0$ galaxies. The black dotted line is consistent with the star formation rate of \CosmosI\ if we assume
its observed \Lya\ line flux has an escape fraction of 100\%, while the red dash-dotted line is a larger star-formation
rate where we assume only 46\% of the \Lya\ line has escaped, the upper limit discussed in \sect{metal}. Thus, \CosmosI\ is also consistent with the FMR. The 
horizontal dotted line is at $1/6\Zsun$, which is the metallicity of the most metal poor high-redshift galaxy yet observed \citep{Erb10}. \vspace{1mm}}
\label{figMassMet}
\end{figure*}

Previous work (McLinden \etal\ in prep. 2013) has determined the best-fit stellar mass for \CosmosI\ to be $1.74^{+0.45}_{-0.36} \times 10^{9}\mbox{ } 
\Msun$ with 1-$\sigma$ uncertainties. This was determined using an analysis of the mass 
distribution from Monte Carlo best fits of the spectral energy distribution (SED) of \CosmosI. These SED fits consider 10 different filter observations, and are thus 
robust to mass, as it normalizes the flux across the whole spectrum. We compare the mass and metallicity constraints of \CosmosI\ with the mass-metallicity relation 
of \citet{Tremonti04} and \citet{Erb06}.
In \fig{figMassMet}, taken from \citet{Finkelstein11}, we plot \CosmosI\ (black square) 
along with the $z\sim 0.1$ SDSS sample of \citet{Tremonti04} (gray points), the $z\sim 2.3$ star-forming galaxies in \citet{Erb06} (green triangles with a fit given by a green dashed line), and the two 
LAEs (z=2.29 and 2.49) of \citet{Finkelstein11} whose metallicities were constrained (red and blue 
circle). The metallicity constraints on \CosmosI\ are not surprising, given its redshift, apparent high ionization parameter, and stellar population and age, and are consistent with the fundamental metallicity relation
(FMR; Mannucci \etal\ 2010) for $z\sim0$ galaxies, where the star-formation rate was determined following 
\citet{Hu99}, with $\dot{M} = 1 \Msun \mbox { yr}^{-1} \times L_{Ly\alpha} / 10^{42}\mbox{ erg s}^{-1}$. We determine the \Lya\ 
line luminosity by taking the \Lya\ line flux from \citet{McLinden11} and given below, combined 
with a luminosity distance of $d_{\rm L}(z$$=$$3.11) = 26.9$ Gpc. Using the observed \Lya\ line flux the 
fundamental metallicity relation is given by the black dotted line.

We now look at the [OIII]/H$\beta$ diagnostic modeled in this work and of \citet{Levesque10}. As discussed before, there is a maximum value 
attained by [OIII]/H$\beta$ (e.g., \fig{figOIII_Hb_vs_NeIII_OII}), where at larger metallicity the gas can cool efficiently and reduce the amount of ionized oxygen, and at lower metallicity there is a lower
oxygen abundance. This maximum is larger for larger ionization parameter, however its effect asymptotes above a few $\times 10^9$ cm s$^{-1}$.
The high ionization parameter galaxies explored in this work
fall in or just above this regime. 
Our models suggest the upper-limit on log[OIII]/H$\beta$ is at or around 0.8, which is consistent with observations (e.g., Fosbury \etal\ 2003; Jaskot \& Oey 2013; Xia \etal\ in prep. 2013). 
To be conservative, we assume an upper value for log[OIII]/H$\beta$ to be 0.95, 2-$\sigma$ above the largest ratio observed in all of the galaxies with large ionization parameter. 
Thus, given the observed [OIII] line fluxes this 
yields 1-$\sigma$ lower limits on the H$\beta$ flux. These models suggest $f(H\beta) >\ 3.4$\Flux for  \CosmosI.

Assuming case B recombination, the ratio between \Lya\ and H$\beta$ is 24.88, following 
\citet{Osterbrock89}. Thus, bounds on H$\beta$ correspond to limits on the emitted, 
pre-escape \Lya. We combine this information with the observed \Lya\ flux (36.14${^{+2.35}_{-2.47}}$\Flux\ for \CosmosI\
from McLinden \etal\ 2011) to get a range of escape fractions of \Lya, $f_{esc,Ly\alpha}$. This suggests that $ f_{\rm esc,Ly\alpha} <$ 46
\% for \CosmosI, consistent with the values found in \citet{Blanc11} for 2 $< z <$ 4 LAEs.  If we consider that only 
46\% of the \Lya\ line is escaping, then the fundamental metallicity relation becomes the 
red dash-dotted in \fig{figMassMet}. 

\subsection{\CosmosII}
Like \CosmosI, \CosmosII\ was found to have a continuous star formation history in McLinden et al. in prep. (2013). The new models yield no further constraints on the metallicity of \CosmosII, as its [OIII]/[OII] ratio implies an ionization parameter in the regime of \citet{Levesque10}. The work of \citet{Levesque10} constrains the ionization parameter of \CosmosII\ to be at least $4\times 10^7\cms$. Considering the lower limits
on H$\beta$ discussed above, we determine  $f(H\beta) >\ 0.6$ \Flux\ for \CosmosII. From \citet{McLinden11}, we have a \Lya\ flux of 
9.41$^{+1.42}_{-1.63}$\Flux\ suggesting an escape fractions of \Lya\ of $ f_{\rm esc,Ly\alpha} <$ 74\%. This is also consistent with the values found in \citet{Blanc11} for 2 $< z <$ 4 LAEs.

\section{Conclusions} \label{conclusion}
We have performed a large suite of models of photoionized regions with large ionization parameters ($q = 4, 6, 9, 15, 30, 45, 60, 90, 150,$ and $300 \times 10^8$ $\mbox{cm s}^{-1}$), low metallicities ($Z = 1, 0.4, 0.2,$ and $0.05\Zdot$), and multiple star formation histories and ages to continue the
work of \citet{Levesque10}. Exploring even larger ionization parameters is likely unwarranted as its effect asymptotes at our most extreme
values.
These models were presented in multiple combinations of line diagnostics,
then applied to our observations of LAEs. We also compare these with recent observations
of high ionization parameter galaxies in Appendix A, finding metallicities and ionization parameters consistent with 
the original presented results. Using these models to constrain the physical
parameters is aided, or in some cases made possible, by knowing the star formation history and age of the galaxy \emph{a priori}, as some diagnostics can have significant variation from one age to the next. 

We emphasize the use of the
[OIII]/[OII] ionization parameter diagnostic with the [NII]/[OII] metallicity diagnostic as it yields unambiguously
both the ionization parameter and metallicity. The [NII]/H$\alpha$ is also a very useful replacement for 
[NII]/[OII] as a metallicity diagnostic since it is insensitive to reddening. Likewise, [OIII]/[OII] can be replaced by 
[NeIII]/[OII] as an ionization parameter diagnostic that is insensitive to reddening, and this is discussed more
in Levesque \& Richardson (2013).

It is important to note that, as demonstrated in multiple diagnostic figures, the model grids presented here cannot fully accommodate 
the observations of [OIII]/H$\beta$, R$_{23}$, and [NeIII]/[OII]. The synthetic ionizing spectra are too soft in the FUV, leading to insufficient line fluxes. This 
was previously discussed by Kewley et al. (2001) and Levesque et al. (2010); many of the models used to calibrate current star-forming 
galaxy diagnostics have similar shortcomings. Updates to stellar population synthesis and photoionization models should therefore 
lead to improved diagnostic calibrations. For example, population synthesis models that include stellar rotation produce harder ionizing 
spectra (Levesque et al. 2012), which in turn should lead to stronger synthetic emission line fluxes and improve the agreement between 
models and observations.

We have observed two Lyman-$\alpha$ emitting galaxies that we fit using these models. We have 
identified line fluxes of [OIII] and [OII] and determined observational 3-$\sigma$ upper limits on [NeIII] for one
$z=3.1$ LAE, and we have identified line fluxes of [OII] and determined observational 3-$\sigma$ upper limits on [NeIII], 
[OII] and H$\beta$ for another
$z=3.1$ LAE.
Future work refining the [OII], [NeIII], and H$\beta$ flux values will better constrain these results, as 
well as provide more insight into the evolution of LAE. We add to the scarce metallicity measurements of line-selected Ly$\alpha$ galaxies with this work.
Specifically:

\CosmosI\ was observed with a line flux of $32.33\pm1.74$ and $11.90 \pm 0.67$\Flux for [OIII] 
5008.240 \AA\ and 4960.295 \AA, respectively. [OII] was detected at the 2$\sigma$ confidence level
as $3.05\pm1.63$\Flux. 3-$\sigma$ upper limits of $1.81$\Flux were determined for [NeIII]. Together these require a metallicity of $Z \leq \Zdot$, consistent with 
\citet{Erb06} and \citet{Xia12}, and either an ionization parameter, $q>2\times 10^8\cms$, making it more ionizing than average starbursting galaxies in the 
local Universe \citep{Rigby04}, or even more metal-poor, with $Z \le 0.2\Zdot$. Upper limits on the [OIII]/H$\beta$ diagnostic were taken which result in an upper 
limit on the escape fraction of \Lya\ photons, $f_{\rm esc,Ly\alpha} < 46$\%, which is consistent with other 
LAEs in this redshift range \citep{Blanc11}. 

\CosmosII\ was observed with a line flux of $6.85\pm1.06$ and $3.56 \pm 0.51$\Flux for [OIII] 5008.240 
\AA\ and 4960.295 \AA, respectively. We have determined 3-$\sigma$ upper limits of $2.92$, $2.57$, and 
$4.34$\Flux for [OII], [NeIII], and H$\beta$, respectively. Together these suggest an ionization parameter, $q>4\times 
10^7 \cms$ \citep{Levesque10}. Upper limits on the [OIII]/H$\beta$ diagnostic were 
taken which result in an upper limit on the escape fraction of \Lya\ photons, $f_{\rm esc,Ly\alpha} < 74$\%. 
Again, this constraint on escape fraction of \Lya\ photons is consistent with \citep{Blanc11}

In summary, we have determined metallicity constraints for a high-redshift LAE, and find such objects to
have large ionization parameters, consistent with primordial galaxies undergoing an early episode of star formation. We have also constrained the escape 
fraction of \Lya\ photons, revealing the effect of dust on \Lya. Together, these results are important for future surveys for 
progressively fainter and more distant LAEs. We urge future work to use our models and those of \citet{Levesque10} for they consider multiple star formation histories and large ionization parameters.

\

The authors would like to thank the the Gemini observing facility and its staff for very useful 
correspondence and discussion during this work. We would like to recognize the cultural 
importance of Mauna Kea with the indigenous population of Hawai'i, and thank them for the 
opportunity to conduct observations from this mountain. This work was supported by NSF grant 
AST-0808165. M. L. A. R was also supported by the National Science and Engineering Research Council of Canada.


\appendix
In this appendix we discuss more of our diagnostic diagrams and how they apply to other high-ionization parameter galaxies discussed in recent literature. We compare and contrast the conclusions
drawn in the original works with those of our models, and discuss possible short-comings in our models. \vspace{3mm}

\section{Additional Line Diagnostics} \label{Appendix}

\subsection{[NeIII]$\lambda 3869$/[OII]$\lambda3727$}
The [NeIII]$\lambda 3869$/[OII]$\lambda3727$ line ratio is described in some detail in \citet{Nagao06}. 
However, they did not explore the dependence of this ratio on ionization parameter. We find that it behaves very much like the [OIII]$\lambda 5007$/[OII] $\lambda 3727$ line ratio (see \fig{figOIII_OII_vs_NeIII_OII} in Appendix \ref{Appendix}), and is best employed as an ionization parameter diagnostic. The dependence 
of ionization parameter on metallicity is responsible for this relation, and is discussed in more detail in
Levesque \& Richardson (2013). Again, it is
weakly dependent on abundance, with larger values corresponding with larger ionization parameter, or lower 
abundance (see for example Figures \ref{figOIII_Hb_vs_NeIII_OII} and \ref{figNII_OII_vs_NeIII_OII}). Of special note, these lines are very similar in wavelength, thus the ratio will not suffer from 
significant reddening effects.

\subsection{[NII]$\lambda 6584$/H$\alpha$}
\begin{figure*}
\centering\leavevmode
\includegraphics[width={0.98\columnwidth}]{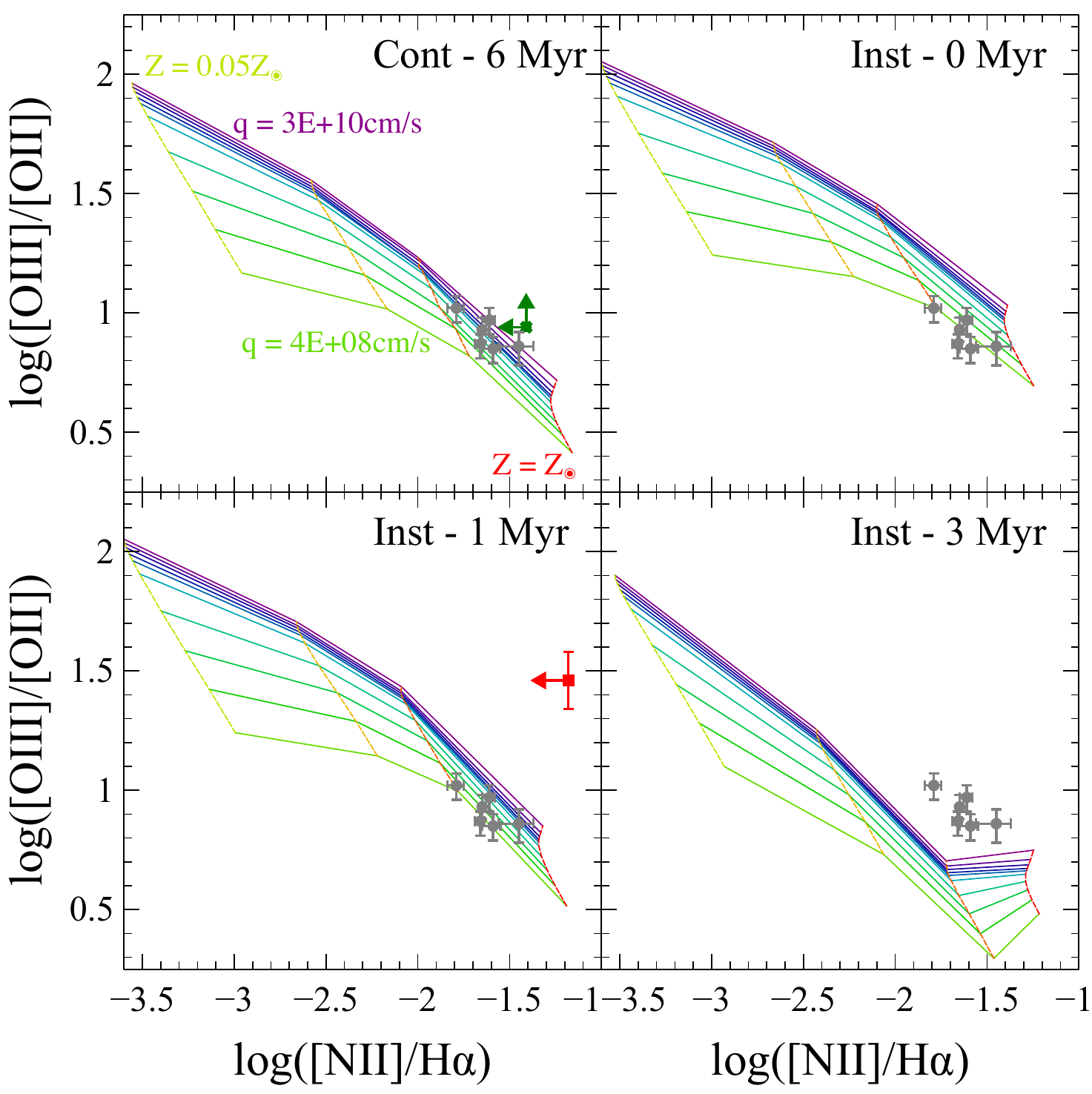} 
\caption{\footnotesize{Similar to \fig{figOIII_OII_vs_NII_OII}, we present here the [OIII]$\lambda 5007$/[OII] $\lambda 3727$ values versus the [NII] 
$\lambda 6594$/H$\alpha$ values. Included are the constraints of Q2343-BX418 by \citet{Erb10} (green flat cross). This object is consistent with larger ionization parameters, noting that the small and intermediate ionization parameter models of \citet{Levesque10} would lie below our models. Q2343-BX418 is also consistent with sub-solar metallicity. We include the constraints and 1-$\sigma$ error bars 
of CDFS-3865 from \citet{Nakajima13} (red square), which is consistent with larger ionization parameters and sub-solar metallicity. Also included are the extreme green peas of \citet{Jaskot13}
 (gray circles), here consistent with near solar metallicity, and mid to large ionization parameter. \vspace{3mm}}}
\label{figOIII_OII_vs_NII_Ha}
\end{figure*}

\begin{figure*}
\centering\leavevmode
\includegraphics[width={0.98\columnwidth}]{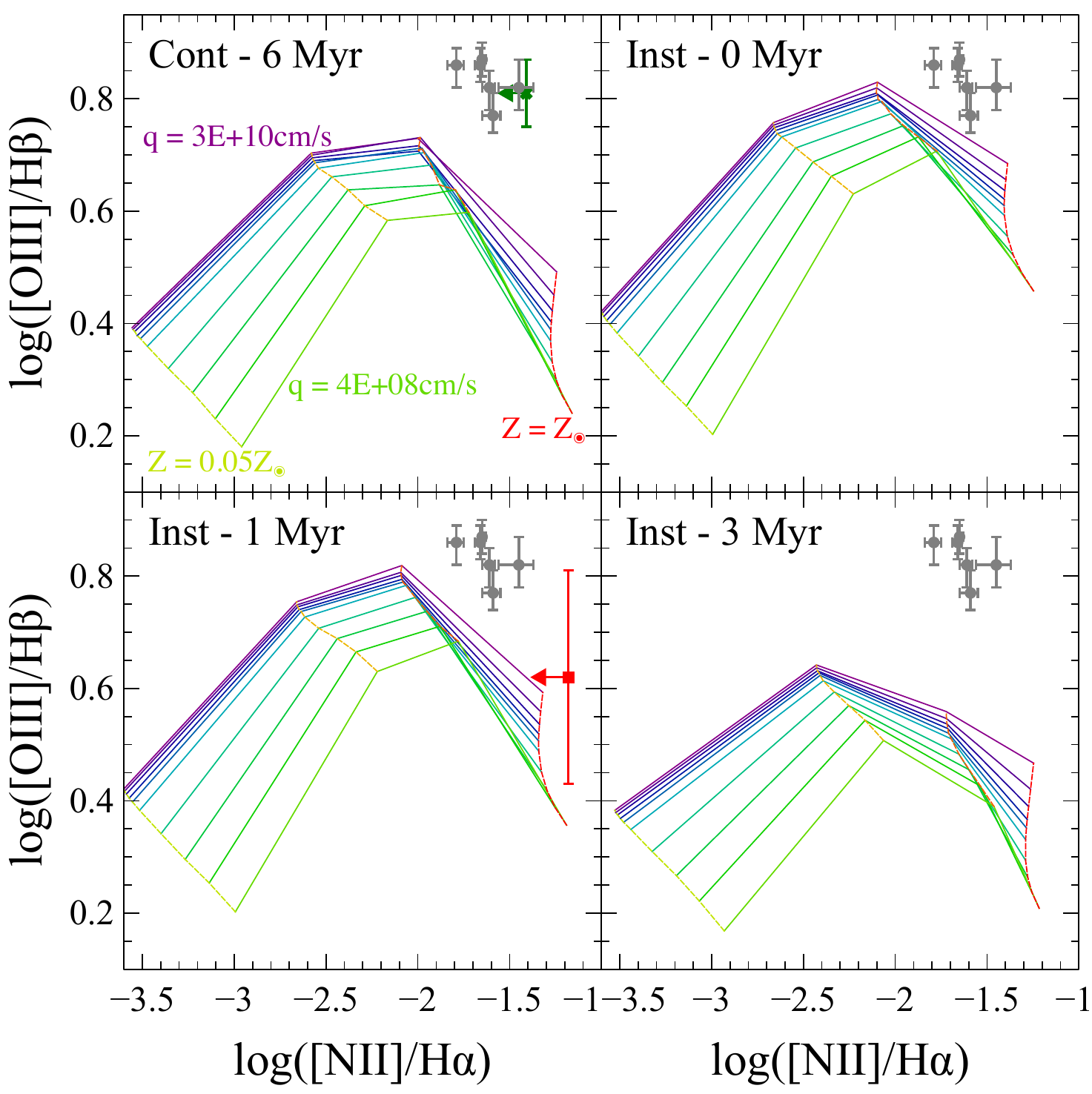} 
\caption{\footnotesize{Similar to \fig{figOIII_OII_vs_NII_OII}, we present here the [OIII]$\lambda 5007$/H$\beta$ values versus the [NII] 
$\lambda 6594$/H$\alpha$ values. Both line diagnostics in this plot are 
insensitive to reddening. Included are the constraints of Q2343-BX418 by \citet{Erb10} (green flat cross). This object is consistent with larger ionization parameters, noting that the small and intermediate ionization parameter models of \citet{Levesque10} would lie below our models. Q2343-BX418 is also consistent with sub-solar metallicity. We Include the constraints and 1-$\sigma$ error bars of
CDFS-3865 by \citet{Nakajima13} (red square), allowing for a wide range of ionization parameter and near solar or lower abundance. Also included are the extreme green peas of \citet{Jaskot13}
 (gray circles), here consistent with near solar metallicity, and possibly very large ionization parameter.\vspace{3mm}}}
\label{figOIII_Hb_vs_NII_Ha}
\end{figure*}

The [NII]$\lambda 6584$/H$\alpha$ line ratio is described in detail in \citet{Levesque10}. It correlates with both metallicity and 
ionization parameter, and is therefore useful in combination with other diagnostics. This ratio is single-valued, with larger values corresponding with larger
ionization parameter, and larger abundances (see for example Figures \ref{figOIII_OII_vs_NII_Ha}, \ref{figOIII_Hb_vs_NII_Ha}, \ref{figNeIII_OII_vs_NII_Ha}, and \ref{figNII_OII_vs_NII_Ha}). Of special note, these lines are very similar in wavelength, thus the ratio will not suffer from 
significant reddening effects.

\subsection{[NII]$\lambda 6584$/[OII]$\lambda 3727$}
The [NII]$\lambda 6584$/[OII]$\lambda 3727$ line ratio is described in detail in \citet{Levesque10}. Since the two lines have
very similar ionization potential, their ratio is almost solely dependent on metallicity, making it an excellent abundance diagnostic. By combining this 
with another line diagnostic sensitive to ionization parameter, like [OIII]$\lambda 5007$/[OII] $\lambda 3727$ or [NeIII]$\lambda 3869$/[OII]$\lambda3727$, one is more capable
of determining both parameters, simultaneously (see for example Figures \ref{figOIII_OII_vs_NII_OII}, \ref{figNII_OII_vs_NeIII_OII}, and \ref{figNII_OII_vs_NII_Ha}). Note that there are inherent assumptions about how this ratio should change with metallicity. At higher metallicities (Z$>$0.4$\Zdot$), [NII] scales more strongly with increasing metallicity, as it is a secondary element. At lower metallicities
this ratio should become more constant as both are primary nucleosynthesis elements. We also stress that this ratio will be particularly susceptible to reddening.

\begin{figure*}
\centering\leavevmode
\includegraphics[width={0.98\columnwidth}]{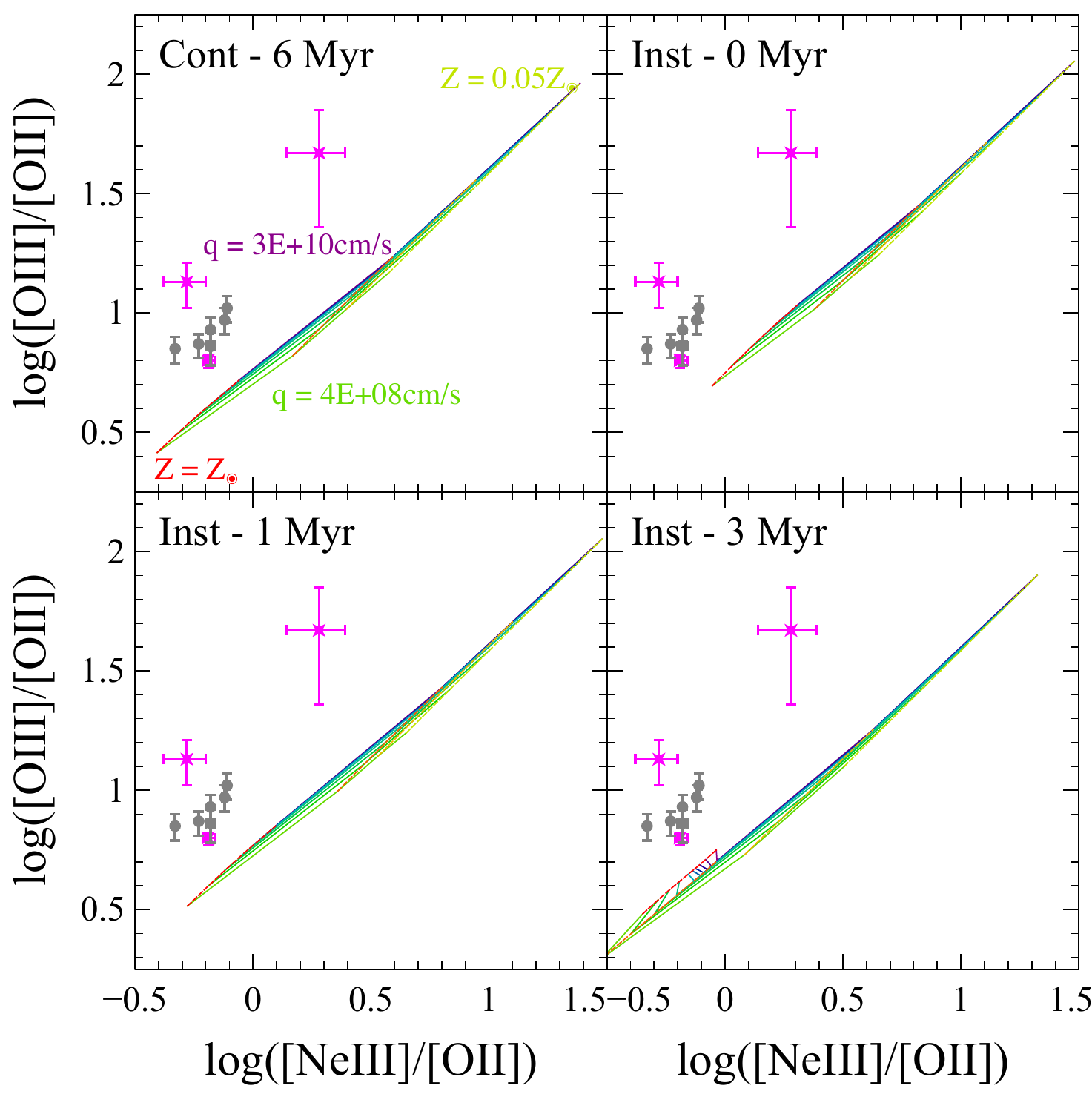} 
\caption{\footnotesize{Similar to \fig{figOIII_OII_vs_NII_OII}, we present here the [OIII]$\lambda 5007$/[OII] $\lambda 3727$ values versus the [NeIII] 
$\lambda 3869$/[OII] $\lambda 3727$ values. The dependence on metallicity and ionization parameter
is incomprehensible as both line ratios are very similar, and much more dependent on ionization potential. This
demonstrates the usefulness of the [NeIII]/[OII] line diagnostic as it is an excellent proxy of [OIII]/[OII] with the added
benefit of being independent of reddening. Further, this figure could be used to demonstrate shortcoming in the 
models in the case where an object does not fall in this grid. Shown are the extreme green peas of \citet{Jaskot13} (gray circles) and objects 20201, 31362, and 823LZ of Xia \etal\ in prep. (2013),
which all lay above the grids.
Work exploring the [NeIII]/[OII] diagnostic as a substitute for the [OIII]/[OII] diagnostic is presented in Levesque \& Richardson (2013). \vspace{3mm}}}
\label{figOIII_OII_vs_NeIII_OII}
\end{figure*}

\begin{figure*}
\centering\leavevmode
\includegraphics[width={0.98\columnwidth}]{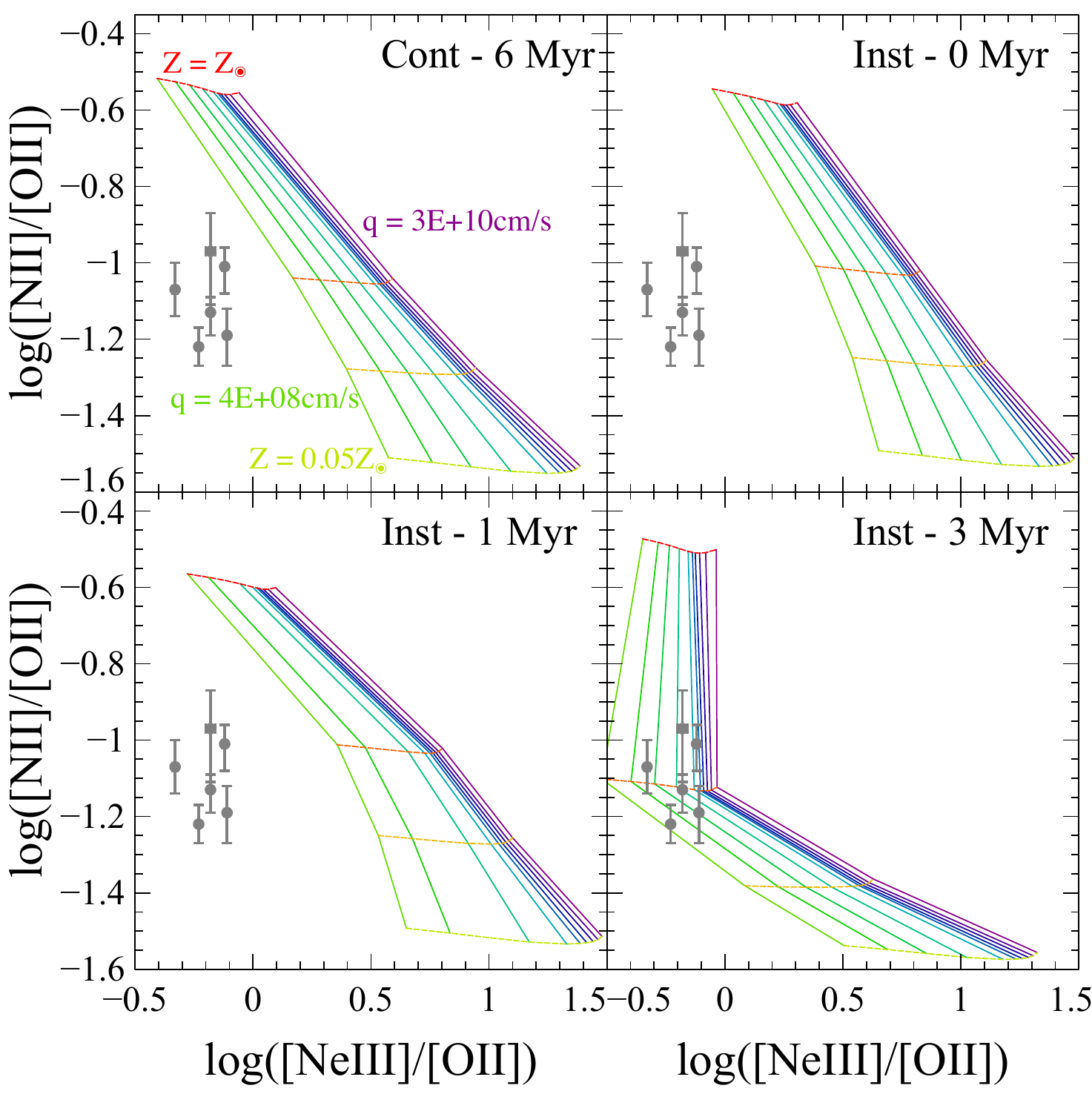} 
\caption{\footnotesize{Similar to \fig{figOIII_OII_vs_NII_OII}, we present here the [NII]$\lambda 6594$/[OII] $\lambda 3727$ values versus the [NeIII] 
$\lambda 3869$/[OII] $\lambda 3727$ values. Shown are the extreme green peas of \citet{Jaskot13} (gray circles)  which are consistent with a young, low ionization
parameter, or an older population with a larger ionization parameter. The metallicities are consistent with $0.4\Zsun$.  \vspace{3mm}}}
\label{figNII_OII_vs_NeIII_OII}
\end{figure*}

\begin{figure*}
\centering\leavevmode
\includegraphics[width={0.98\columnwidth}]{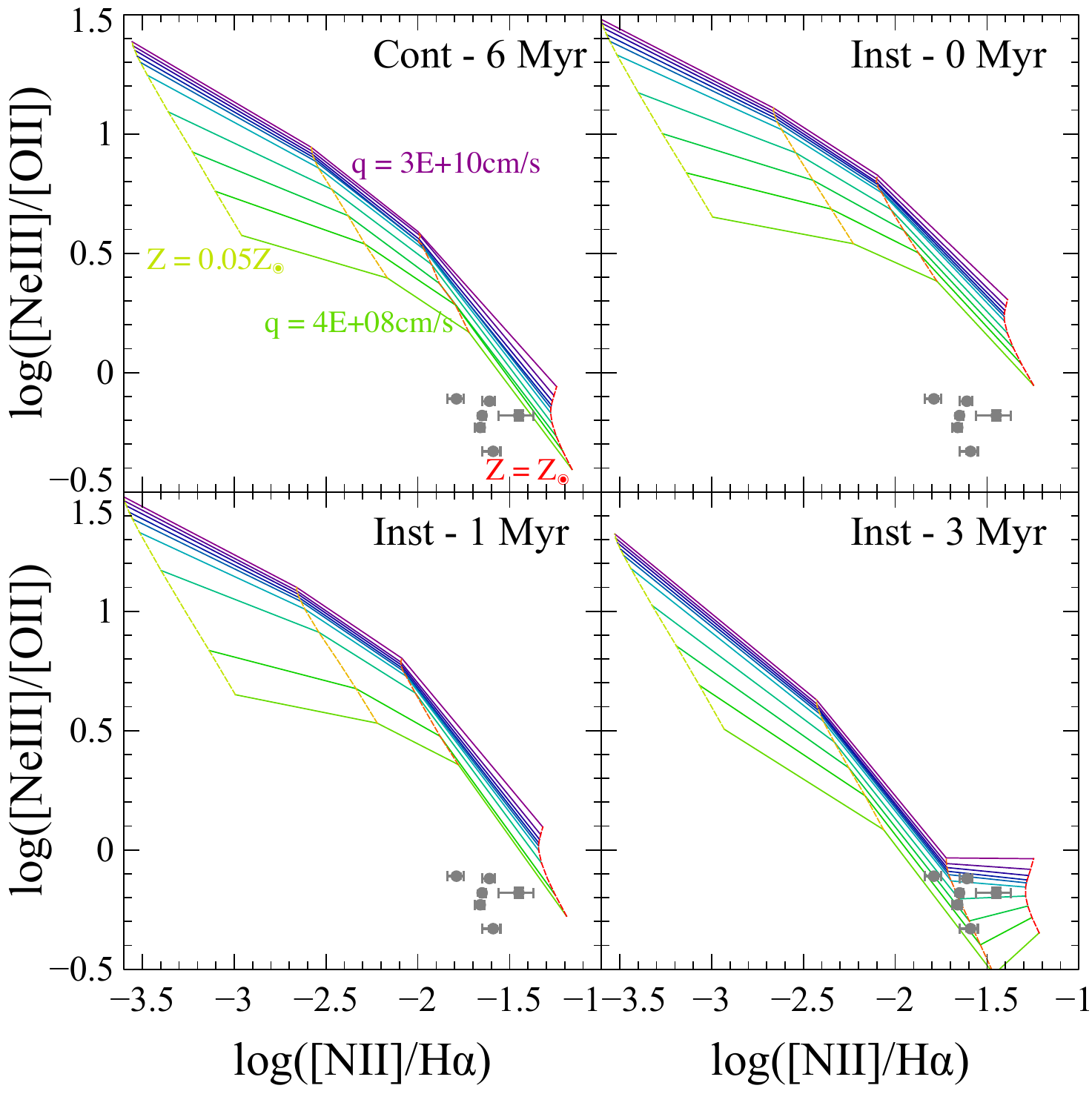} 
\caption{\footnotesize{Similar to \fig{figOIII_OII_vs_NII_OII}, we present here the [NeIII]$\lambda 3869$/[OII] $\lambda 3727$ values versus the [NII] 
$\lambda 6594$/H$\alpha$ values. Both line diagnostics in this plot are 
insensitive to reddening. Shown are the extreme green peas of \citet{Jaskot13} (gray circles)  which are consistent with a young, low ionization
parameter, or an older population with a larger ionization parameter. The metallicities are consistent with near solar, depending on age.  \vspace{3mm}}}
\label{figNeIII_OII_vs_NII_Ha}
\end{figure*}

\begin{figure*}
\centering\leavevmode
\includegraphics[width={0.98\columnwidth}]{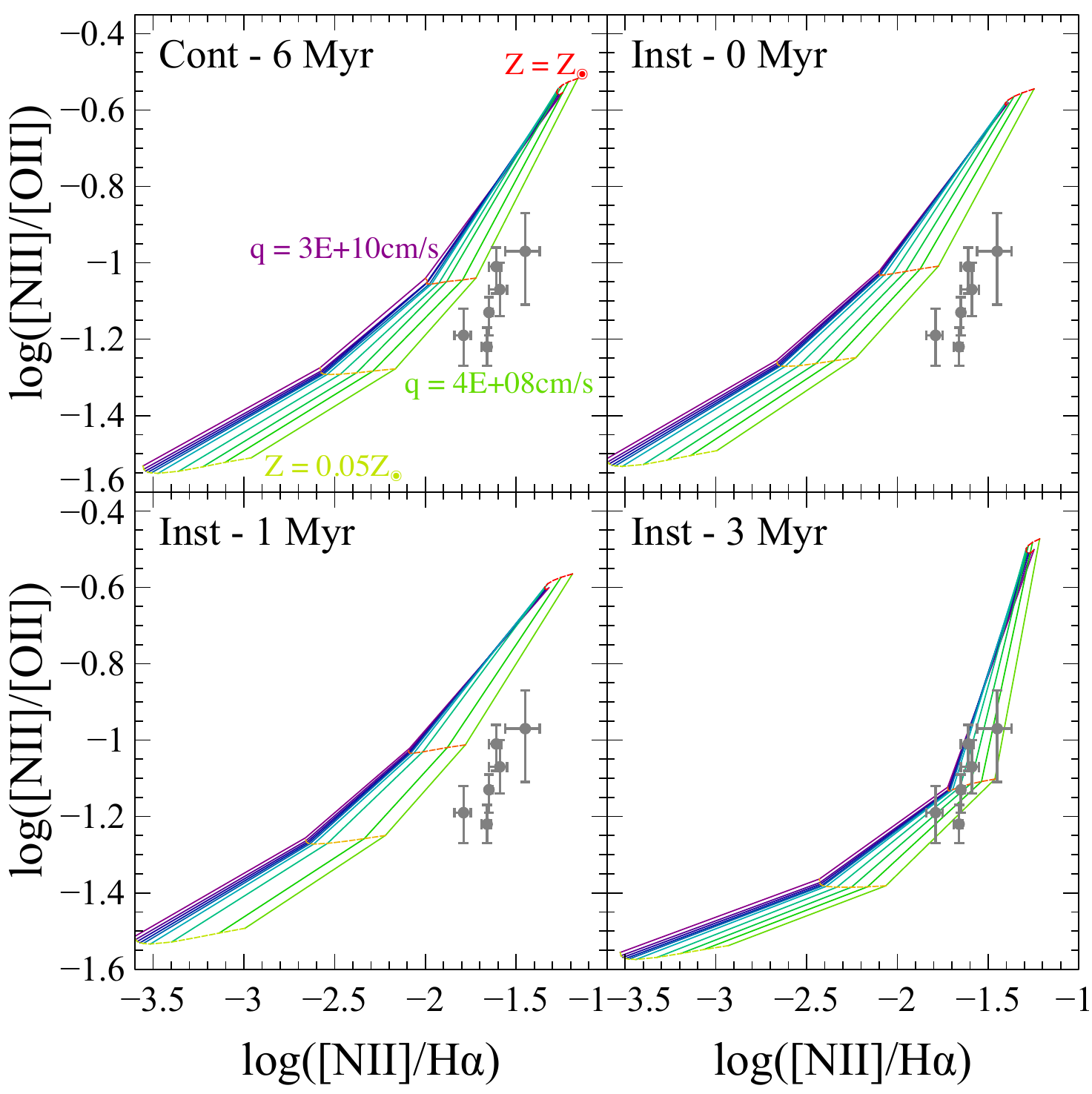} 
\caption{\footnotesize{Similar to \fig{figOIII_OII_vs_NII_OII}, we present here the [NII]$\lambda 6594$/[OII] $\lambda 3727$ values versus the [NII] 
$\lambda 6594$/H$\alpha$ values. Shown are the extreme green peas of \citet{Jaskot13} (gray circles)  which are consistent with a young, low ionization
parameter, or an older population with a larger ionization parameter. The metallicities are consistent with $0.4\Zsun$.  \vspace{3mm}}}
\label{figNII_OII_vs_NII_Ha}
\end{figure*}

\section{Model Comparison with High Ionization Galaxies}
\begin{figure*}
\centering\leavevmode
\includegraphics[width={.98\columnwidth}]{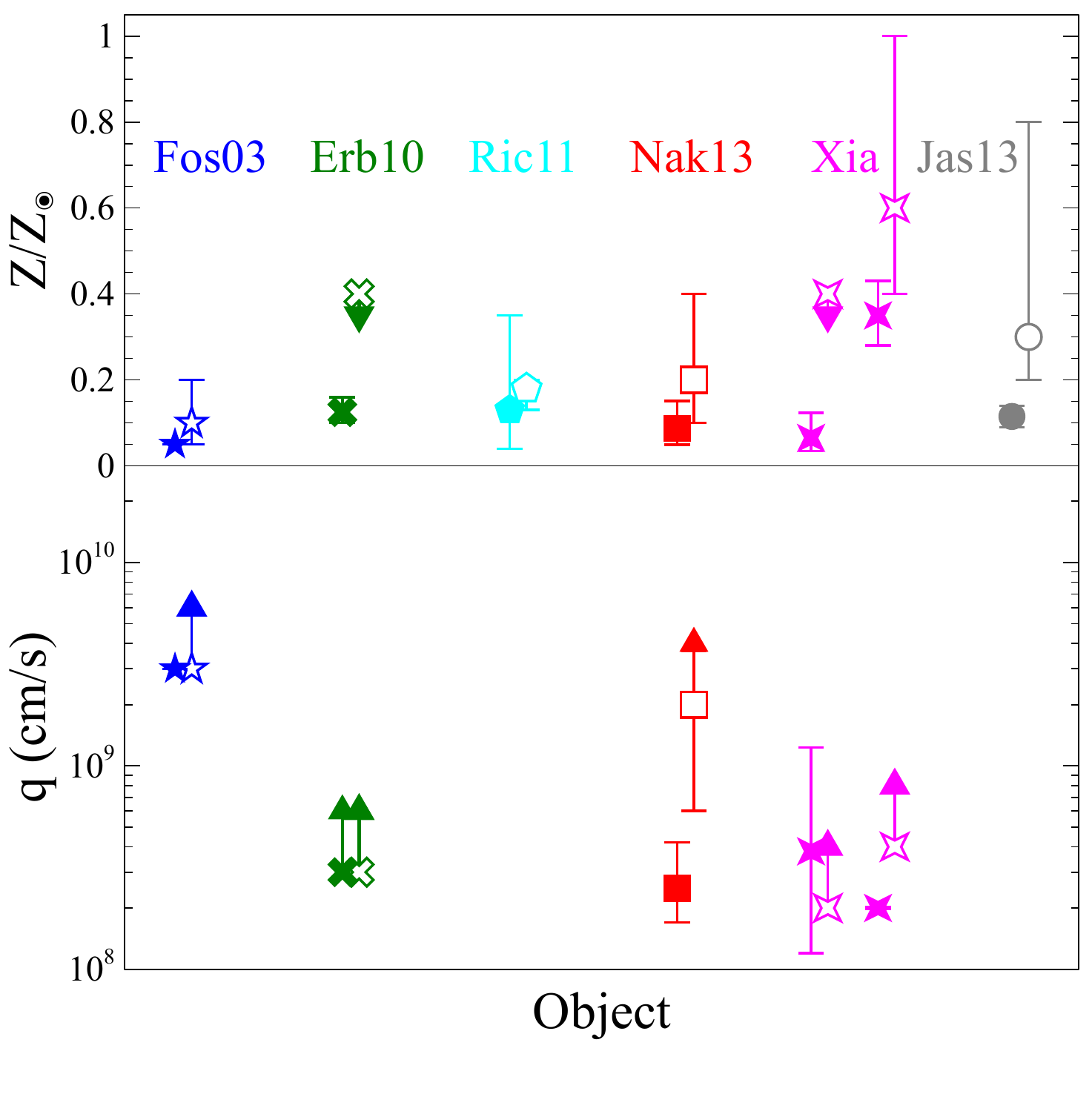} 
\caption{\footnotesize{Shown here are the comparison between the characteristics of high ionization parameters as stated in their original work (filled) and as determined in this work (empty). Colors
and shapes are the same as in Figures 3-12. We compare the metallicity and the ionization parameter (when available). All metallicities are in agreement (to 2-sigma), while one 
ionization parameter is increased.  \vspace{3mm}}}
\label{figLit}
\end{figure*}

We have included observations of \citet{Fosbury03}, \citet{Erb10}, \citet{Richard11}, \citet{Xia12} and Xia \etal\ in prep 2013, and \citet{Jaskot13} in our line 
diagnostic diagrams. Most of these observations include a star-forming history for each object with the exception of  \citet{Xia12} and Xia \etal\ in prep 2013, and \citet{Jaskot13}. 
When we determine what constraints our models place on the metallicity and ionization parameter of the remaining objects, we find values that are
consistent with their original work. \fig{figLit} compares our constraints (empty symbols) with those in their original work (filled). An exception to the
agreement is with the ionization parameter of \citet{Nakajima13}, which is only consistent to 2-$\sigma$. We stress however their use of work that does not 
consider higher ionization parameters.

We have included the objects of \citet{Xia12} and Xia \etal\ in prep 2013, and \citet{Jaskot13} in multiple star forming histories to determine if a particular 
history is consistent in metallicity and ionization parameter. These objects are all $z<1$ emission-line selected galaxies. 
In general we find the [OIII]/H$\beta$ and R$_{23}$ ratios are inconsistent with the data, suggesting that our models do not produce sufficiently hard FUV photons (see \sect{results}),
thus we do not consider these diagnostics when constraining these objects. Looking at the other diagrams, we find they allow two possibilities for the \citet{Jaskot13} objects.
Either the populations are an older instantaneous burst with near solar metallicities and large ionization parameters, or somewhat smaller metallicities with
younger stellar populations and significantly lower ionization parameters. Accepting this range, our models are still consistent to 2-$\sigma$ with the metallicities
quoted in \citet{Jaskot13}. Likewise for the objects of Xia \etal, we find the population is either older, less enriched with large ionization parameters, or younger, more enriched, 
with lower ionization parameters. Our abundances are consistent with theirs while our the discrepancy in ionization parameter is expected as they use older work
that does not consider larger ionization parameters and older ages. 

In summary, we wish to stress the difficulty in determining constraints on the physical parameters of these objects without an indication of the star formation history. Figures 3-5, and 
7-12 illustrate this in the variation of the grids from one star formation history to the next.

\end{document}